\documentclass{emulateapj}

\usepackage{natbib}
\usepackage{epsfig}
\usepackage{graphicx}
\usepackage{subfigure}
\usepackage{float}
\usepackage{amsmath}
\usepackage{color}
\usepackage{amssymb}
\usepackage{amsfonts}
\usepackage[colorlinks,linkcolor=blue,anchorcolor=green,citecolor=blue]{hyperref}
\usepackage{amssymb}
\usepackage{upgreek}
\usepackage[normalem]{ulem}
\usepackage{verbatim}
\usepackage[flushleft,bf]{caption2}

\newcommand{\be}{\begin{equation}}
\newcommand{\ee}{\end{equation}}
\newcommand{\ba}{\begin{eqnarray}}
\newcommand{\ea}{\end{eqnarray}}

\begin{document}

\title{Spectroscopic and Photometric Redshift Estimation by Neural Networks \\For the China Space Station Optical Survey (CSS-OS)}
\author{Xingchen Zhou\altaffilmark{1,2}, Yan Gong\altaffilmark{*1,3}, Xian-Min Meng\altaffilmark{1}, Xin Zhang\altaffilmark{1}, Ye Cao\altaffilmark{1,2}, Xuelei Chen\altaffilmark{4,2,5}, \\Valeria Amaro\altaffilmark{6}, Zuhui Fan\altaffilmark{7,8}, Liping Fu\altaffilmark{9}}
\affil{$^1$Key Laboratory of Space Astronomy and Technology, National Astronomical Observatories, \\Chinese Academy of Sciences, 20A Datun Road, Beijing 100101, China. $\it Email:$ gongyan@bao.ac.cn}
\affil{$^2$University of Chinese Academy of Sciences, Beijing 100049, China}
\affil{$^3$Science Center for China Space Station Telescope, National Astronomical Observatories, \\Chinese Academy of Sciences, 20A Datun Road, Beijing 100101, China}
\affil{$^4$Key Laboratory for Computational Astrophysics, National Astronomical Observatories, \\Chinese Academy of Sciences, 20A Datun Road, Beijing 100101, China}
\affil{$^5$Centre for High Energy Physics, Peking University, Beijing 100871, People’s Republic of China}
\affil{$^6$School of Physics and Astronomy, Sun Yat-sen University, Zhuhai Campus, Guangzhou 519082, PR China}
\affil{$^7$South-Western Institute for Astronomy Research, Yunnan University, Kunming 650500, People’s Republic of China}
\affil{$^8$Department of Astronomy, School of Physics, Peking University, Beijing 100871, People’s Republic of China}
\affil{$^9$Shanghai Key Lab for Astrophysics, Shanghai Normal University, Shanghai 200234, China}

\begin{abstract}
The estimation of spectroscopic and photometric redshifts (spec-$z$ and photo-$z$) is crucial for the future cosmological surveys. It can directly affect several powerful measurements of the Universe, e.g. weak lensing and galaxy clustering. In this work, we explore the accuracies of spec-$z$ and photo-$z$ that can be obtained in the China Space Station Optical Surveys (CSS-OS), which is a next-generation space survey, using neural network. The 1-dimensional Convolutional Neural Networks (1-d CNN) and Multi-Layer Perceptron (MLP, a simplest form of Artificial Neural Network) are employed to derive the spec-$z$ and photo-$z$, respectively. The mock spectral and photometric data used for training and testing the networks are generated based on the COSMOS catalog. The networks have been trained with noisy data by creating Gaussian random realizations to reduce the statistical effects, resulting in similar redshift accuracy for both high-SNR (signal to noise ratio) and low-SNR data. The probability distribution functions (PDFs) of the predicted redshifts are also derived via Gaussian random realizations of the testing data, and then the best-fit redshifts and 1-$\sigma$ errors also can be obtained. We find that our networks can provide excellent redshift estimates with accuracies $\sim$0.001 and 0.01 on spec-$z$ and photo-$z$, respectively. Compared to existing photo-$z$ codes, our MLP has similar accuracy but is more efficient in the training process. The fractions of catastrophic redshifts or outliers can be dramatically suppressed comparing to the ordinary template-fitting method. This indicates that the neural network method is feasible and powerful for  spec-$z$ and photo-$z$ estimations in the future cosmological surveys.
\end{abstract}



\section{Introduction}
	Understanding the properties of dark matter and dark energy, and formation and evolution of the cosmic large-scale structure (LSS) are important tasks for cosmological observations. The ongoing and next-generation ground and space telescopes, such as Dark Energy Spectroscopic Instrument (DESI)\footnote{https://www.desi.lbl.gov/}, the Large Synoptic Survey Telescope\footnote{https://www.lsst.org/} \citep[LSST;][]{Ivezic08,Abell09}, {\it Euclid space telescope}\footnote{https://www.euclid-ec.org/} \citep{Laureijs11}, and Wide Field Infrared Survey Telescope\footnote{https://wfirst.gsfc.nasa.gov/} (WFIRST), are dedicated to answering these questions with ultrahigh precisions. By then billions of galaxies will be detected in enormous spatial volumes over large redshift ranges via spectroscopy and imaging photometry, and kinematic and dynamical evolutions of the Universe can be precisely measured. All of these goals are tightly related to the measurements of galaxy redshifts. Obtaining accurate spectroscopic and photometric redshifts (spec-$z$ and photo-$z$) is strict requirement for these future cosmological surveys.
	
	The China Space Station Telescope (CSST) is a 2-meter space telescope, focusing on the next generation cosmological observations. It currently has been scheduled to launch in 2024, and is in the same orbit of the China Manned Space Station \citep{Zhan11, Zhan18, Cao18, Gong19}. The China Space Station Optical Survey (CSS-OS) is one of the main missions of the CSST, and it is planned to observe 17,500 deg$^2$ in about ten years, covering optical and near-IR bands from $\sim$250 to 1000 nm.  It can simultaneously perform photometric and slitless spectroscopic surveys with large field of view and high spatial resolutions. The scientific goals of the CSS-OS include exploring the properties of dark matter and dark energy, evolution of the LSS, galaxy formation and evolution, galaxy clusters, etc. Several kinds of cosmological and astrophysical observations will be performed by the CSST to achieve these goals, such as weak and strong gravitational lensing, 2- and 3-dimensional (2- and 3-d) galaxy clustering, galaxy and active galactic nucleus (AGN) surveys and so on. 
	
	Weak lensing and galaxy clustering are powerful probes in cosmological studies, which could provide stringent constraints on dark matter, dark energy, and evolution of the LSS. Since the accuracies of spec-$z$ and photo-$z$ can directly affect the measurements of shear correlation function of weak lensing and 2- and 3-d galaxy power spectrum, they are crucial for the CSS-OS. The spec-$z$ and photo-$z$ usually can be derived by fitting galaxy spectral energy distribution (SED) templates. However, this method is highly dependent on the quality of selected SED templates, that the accurate spec-$z$ and photo-$z$ data may not be obtained if the SED templates adopted are not sufficiently representative. Besides, poor spectroscopic and photometric measurements can also lead to inaccurate spec-$z$ and photo-$z$ fitting in this method. This is particularly the case for high-redshift and faint galaxies, slitless spectroscopic surveys with low spectral resolution (like CSST), and photometric bands with small filter quantum transmission efficiency (e.g. CSST $NUV$ and $y$ bands). 

	On the other hand, we can try to find empirical relations between galaxy properties (e.g. SED, band magnitude, color, morphology, etc.) and redshift by exploring a huge amount of galaxies with known redshifts. This can be achieved by training neural networks with appropriate training datasets,  which can be obtained by processing high-quality data from ongoing or future spectroscopic surveys, such as Multi-Object Optical and Near-infrared Spectrograph (MOONS)\footnote{https://www.eso.org/sci/facilities/develop/instruments/\\MOONS.html}\citep{Cirasuolo20,Maiolino20}, Prime Focus Spectrograph (PFS) \citep{Tamura16}, DESI, 4-metre Multi-Object Spectroscopic Telescope (4MOST)\footnote{https://www.4most.eu/cms/}, MegaMapper \citep{Schlegel19}, Fiber-Optic Broadband Optical Spectrograph (FOBOS) \citep{Bundy19}, SpecTel \citep{Ellis19} and so on. These powerful spectroscopic surveys would get large amounts of high-quality galaxy SEDs at $z=0\sim4$, which can be used as training data for the CSST spectroscopic and photometric surveys.
	
	In recent years, development of machine learning is remarkable, especially for neural network algorithm. Neural network is widely applied for astronomy and cosmology studies, especially for the  Multi-Layer Perceptron (MLP) and Convolutional Neural Network (CNN). The MLP is the simplest form of the Artificial Neural Network (ANN), which consists of an input layer, several hidden layers, and an output layer. The CNN is first proposed by \cite{Fukushima70} and \cite{Lecun98}, which is a kind of neural network applied widely nowadays in our daily life and scientific research. CNNs can process data with multiple arrays, such as images with RGB channels, and they have great success in detection, segmentation and recognition applications. In this work, the 1-d CNN and MLP are adopted to derive spec-$z$ and photo-$z$ for the CSS-OS, respectively. The Adam optimizer is adopted to automatically update the network learning rate \citep{Kingma14}, which is more efficient for network training than the previous machine learning codes used in the redshift estimation \citep[e.g.][]{Collister04,Gerdes10,Samui17}.
	
	We make use of the COSMOS galaxy catalog \citep{Capak07,Ilbert09} to generate mock spectroscopic and photometric data for training and testing the neural networks. The probability distribution function (PDF), best-fit (the value at the peak of PDF) and 1-$\sigma$ error of predicted redshift can be obtained in our analysis. In this study, we try to figure out a number of important questions about the neural networks applied to the CSS-OS, e.g. how accurate that the CSS-OS spec-$z$ and photo-$z$ can achieve using this method, how many training data are needed in this process, can the redshifts be correctly derived from low-quality data, and so on. The answers of these questions would be quite helpful to determine the detailed strategy of extracting the spec-$z$ and photo-$z$ using neural network for the real data processing of the CSS-OS in the future.
	
	This paper is organized as follows: In Section~\ref{sec:MOCK DATA}, we discuss the generation of  mock data we used for training and testing. In Section~\ref{sec:Neural Network}, we show the architectures of the adopted neural networks and details of training process for the CSST spec-$z$ and photo-$z$ predictions. The results are presented in Section~\ref{sec:Result}. We summarize our results in Section~\ref{sec:Summary}. 

\section{Mock Data}
\label{sec:MOCK DATA}

\begin{figure}
\centering
\includegraphics[scale=0.45]{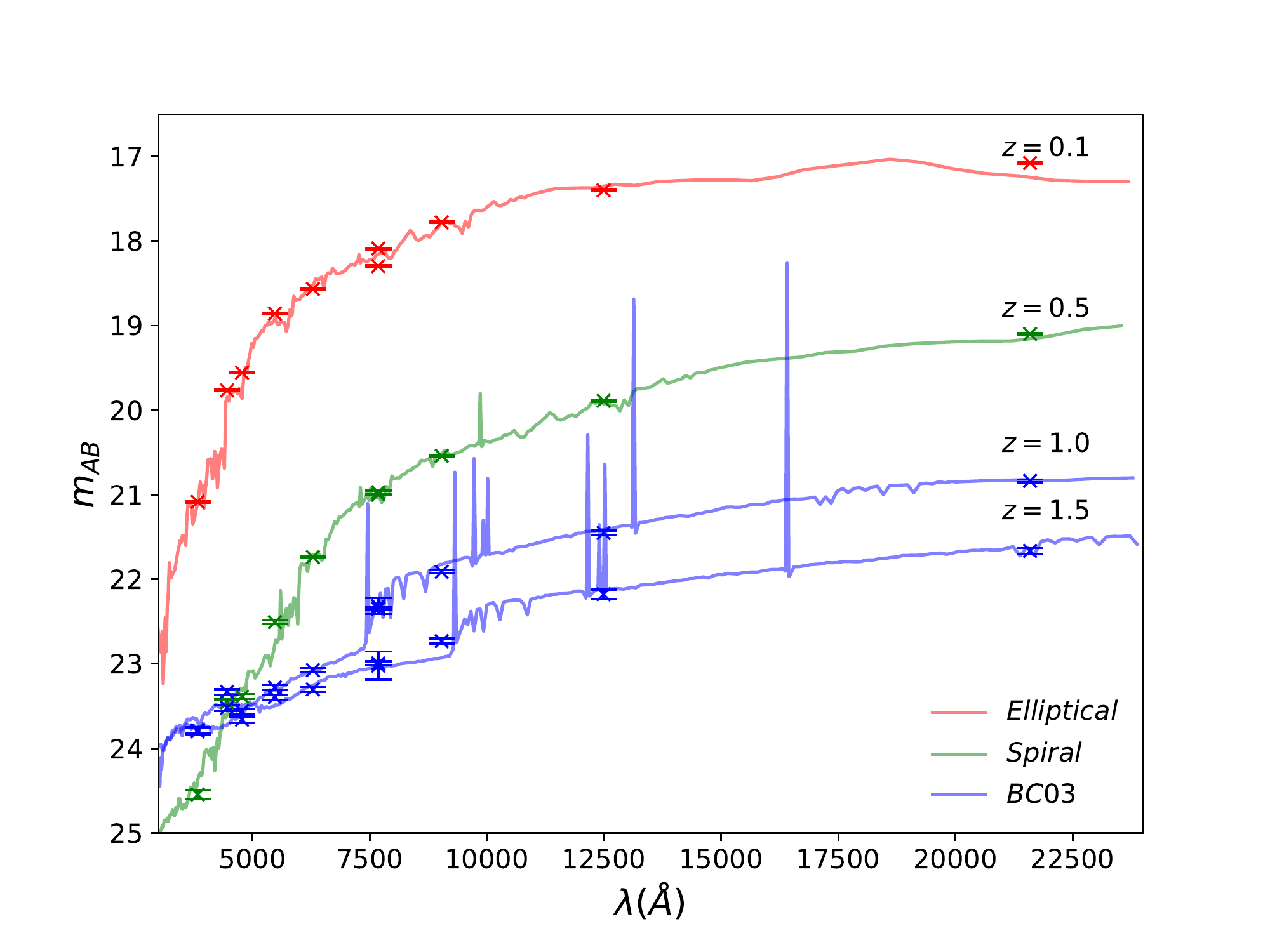}
\caption{Examples of the derived SEDs with emission lines by refitting the photometric data of the COSMOS catalog using the $\it LePhare$ code. For comparison, the photometric data in the ten bands are also shown, i.e. $u^*$ (CFHT), $B_{\rm J}$ (Subaru), $V_{\rm J}$ (Subaru), $g^+$ (Subaru), $r^+$ (Subaru), $i^+$ (Subaru), $i*$ (CFHT), $z^+$ (Subaru), $J$ (UKIRT), and $K$ (CFHT) \citep{Ilbert09}. The red, green and blue curves and data points are for elliptical, spiral, and young blue star-forming galaxies (denoted as BC03, which are fitted by the method given in \cite{Bruzual03}), respectively.}
\label{fig:Spec_fit_examples}
\end{figure}

\begin{figure*}
\centering
\includegraphics[width=3.5in]{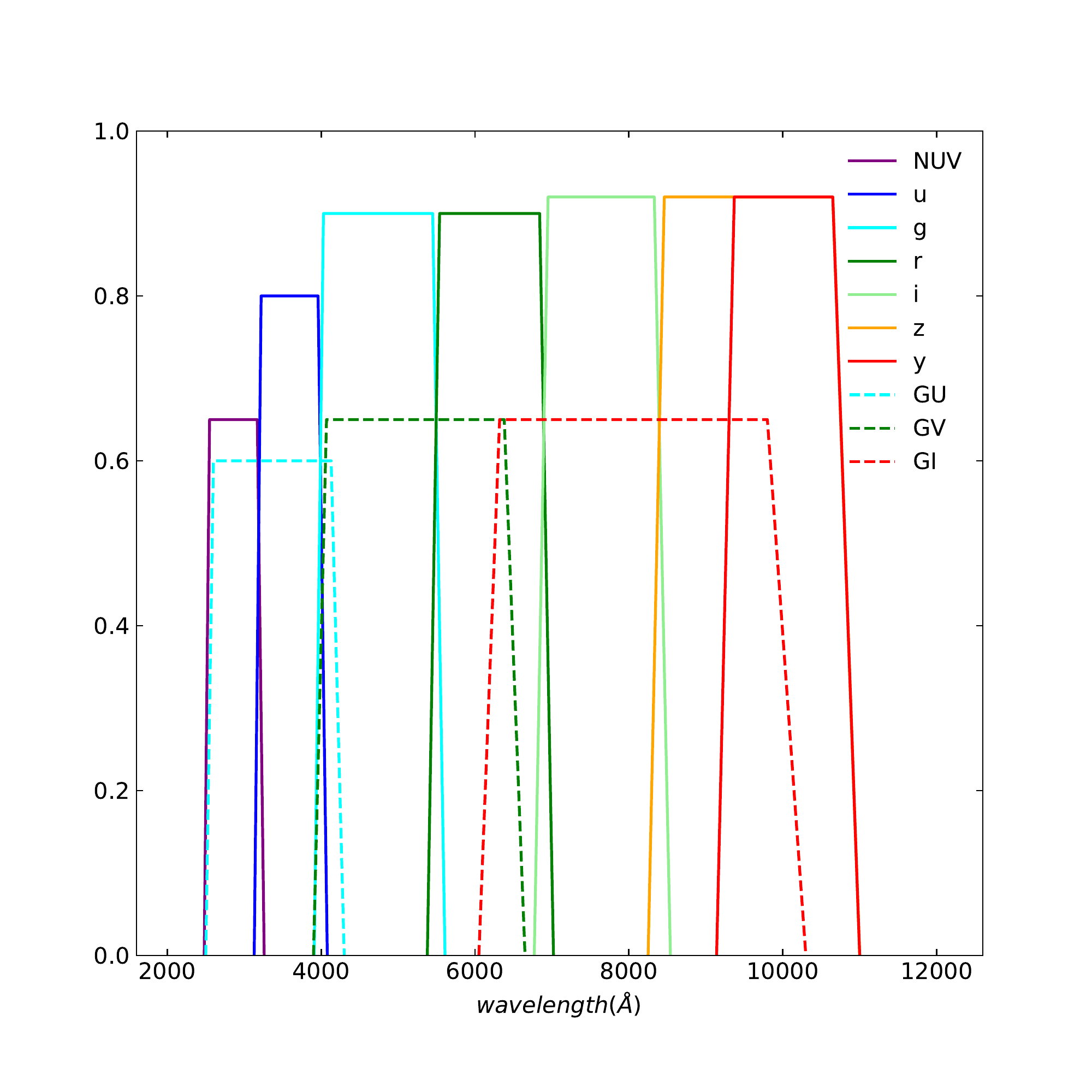}
\includegraphics[width=3.5in]{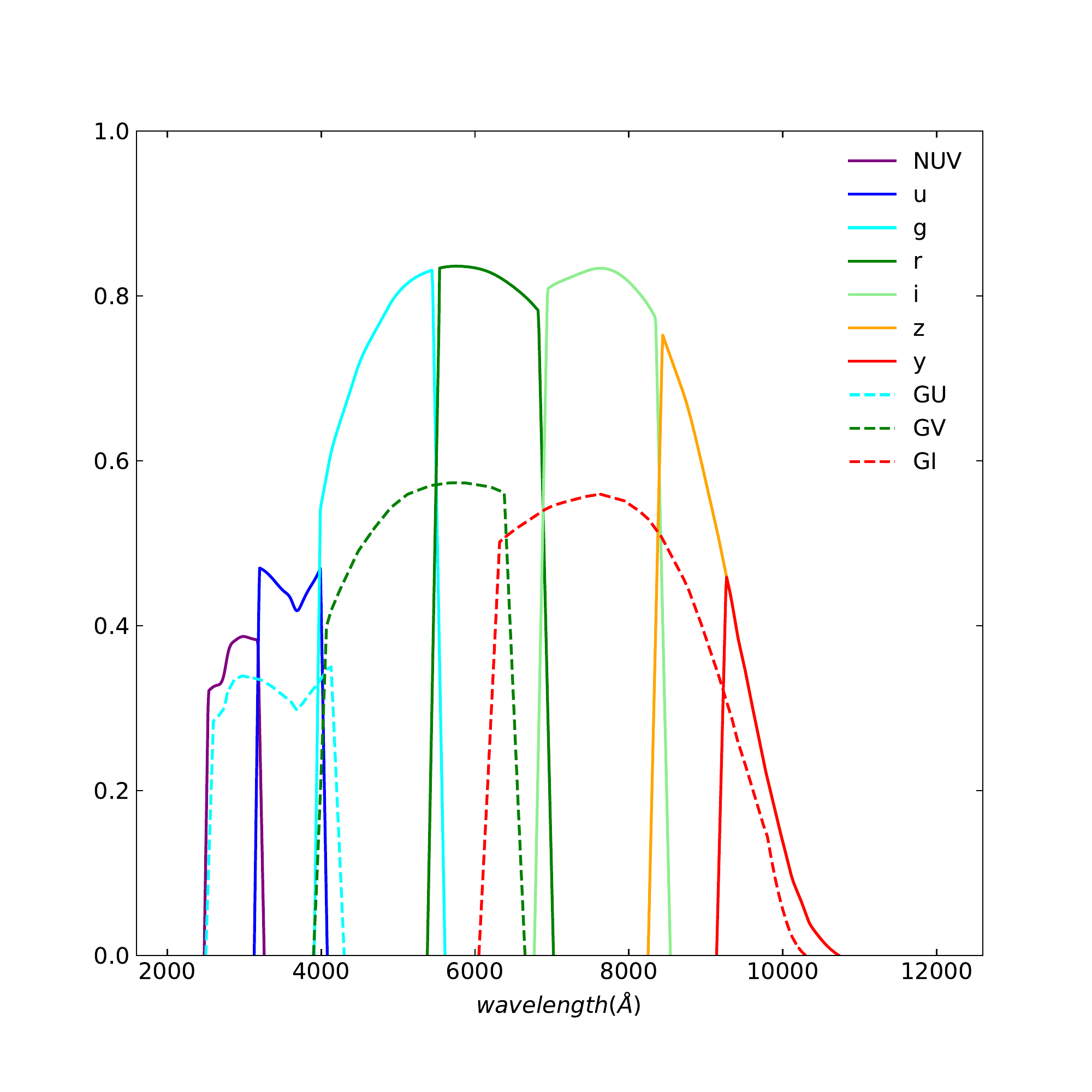}
\caption{\emph{Left panel}: The transmission curves of the CSST seven photometric (solid) and three spectroscopic (dashed) bands, covering from near-UV to near-IR bands \citep{Cao18,Gong19}. \emph{Right panel}: The total transmission curves considering quantum efficiency \citep{Cao18}.}
\label{fig:Transmission}
\end{figure*}

In this section, we discuss the generation of spectroscopic and photometric mock data for the CSS-OS. The mock data need to correctly represent the galaxy observations of the CSS-OS, which should have similar properties, e.g. redshift and magnitude distributions, and galaxy types. Here the COSMOS galaxy catalog is used to create the mock data, which contains about 220,000 galaxies with $i^+\le25.2$ in $\sim$2 deg$^2$ including the information of galaxy redshift, magnitude, size, dust extinction, best-fit SED, and so on \citep{Capak07,Ilbert09,Cao18,Gong19}. Since the galaxy redshifts and best-fit SEDs can be seen as high-quality data obtained by a number of powerful spectroscopic galaxy surveys (as listed in the introduction), with the help of the information given by the catalog, we can generate the training and testing samples for both CSST spectroscopic and photometric surveys by considering the CSST observational and instrumental effects. As mentioned in the following discussion, the main selection criteria of the CSST spectroscopic and photometric mock data are the magnitude limit and signal to noise ratio (SNR).

\begin{figure*}
\centering
\includegraphics[width=3.5in]{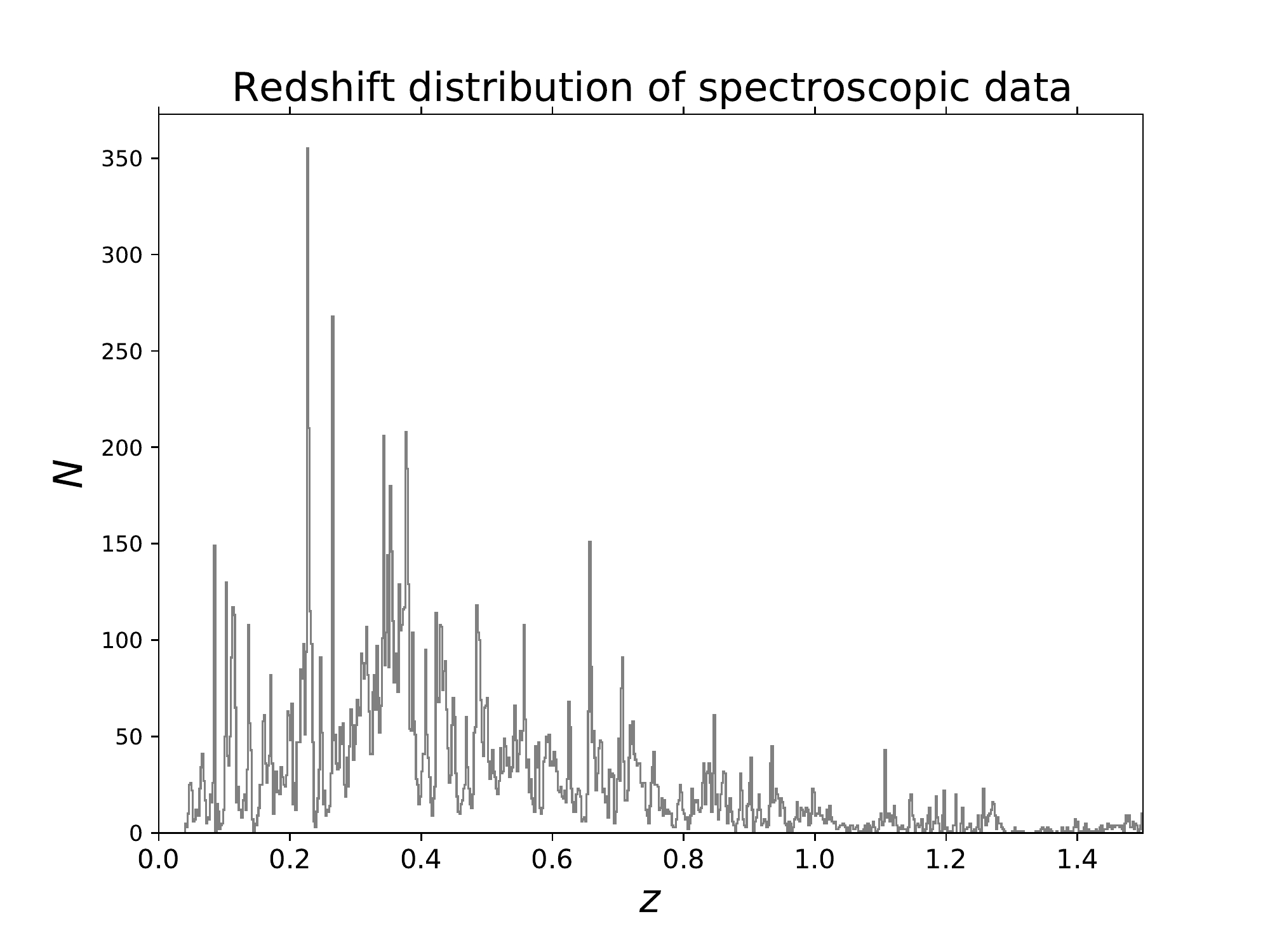}
\includegraphics[width=3.5in]{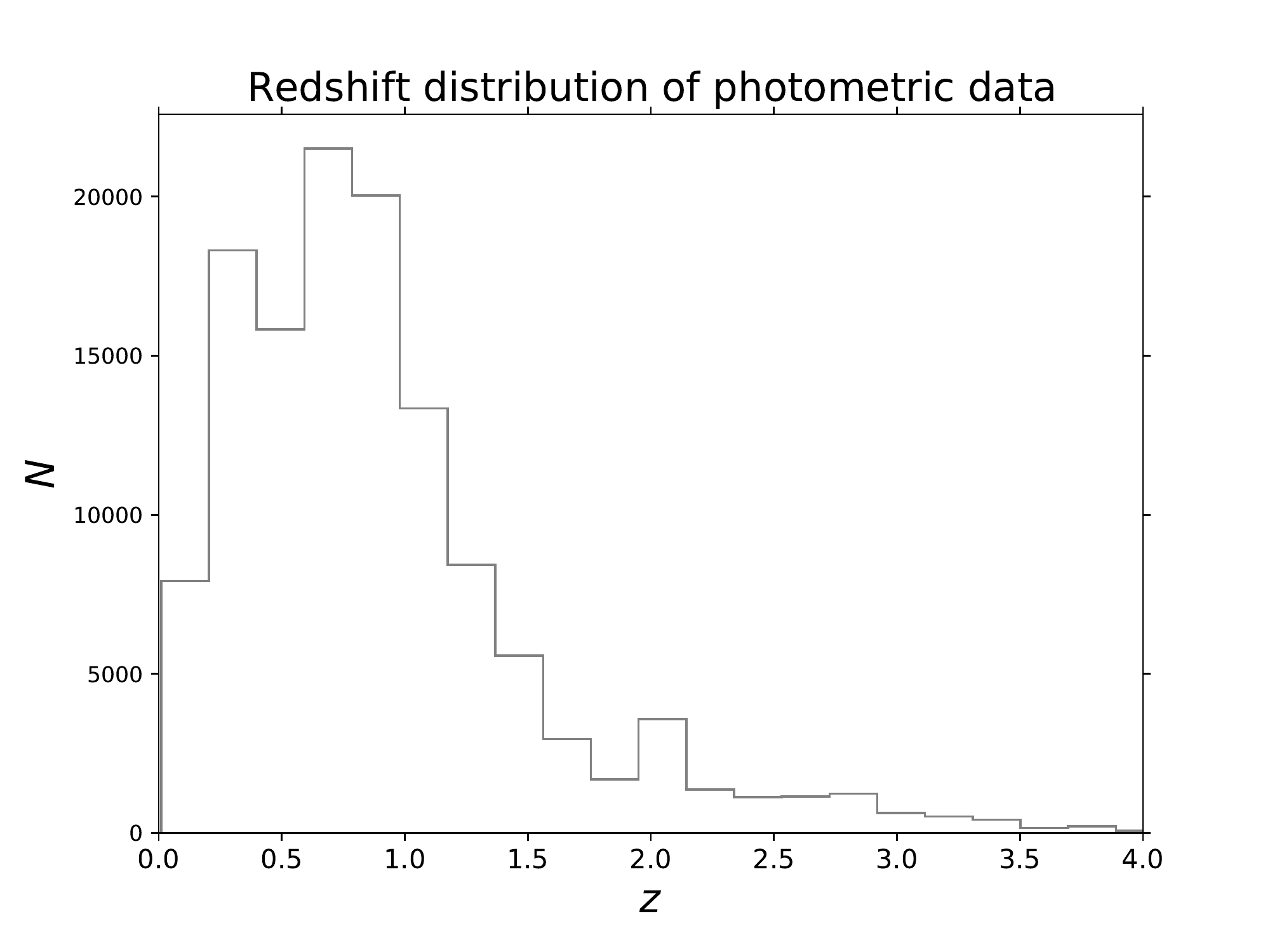}
\caption{The galaxy redshift distributions of the CSST spectroscopic (left panel) and photometric (right panel) surveys derived from the COSMOS catalog. These redshift distributions are consistent with previous studies given by \cite{Cao18} and \cite{Gong19}. The redshift bin size is d$z=0.002$ in the left panel, which matches the d$z$ adopted in the training and testing processes of the CNN as discussed in Sec.~\ref{sec:Neural Network}.}
\label{fig:spec_photo_zdis}
\end{figure*}

\subsection{Spectroscopic mock data}

The CSST will perform spectroscopic survey using slitless gratings. It has three bands, i.e. $GU$, $GV$, and $GI$, as shown in dashed curves in the two panels of Figure~\ref{fig:Transmission} for the transmissions that originally designed and quantum-efficiency considered cases, respectively. The AB magnitude 5$\sigma$ limits of the three bands are 23.1, 23.4, and 23.5 for point sources, respectively, with spectral resolution $R=\lambda/\Delta\lambda\gtrsim200$ \citep{Gong19}. Based on these survey parameters, we can select the galaxy samples and generate the mock data from the COSMOS catalog. 

Firstly, In order to obtain the galaxy SEDs with emission lines, we refit the galaxy photometric data of the COSMOS catalog using the $\it LePhare$\footnote{https://www.cfht.hawaii.edu/~arnouts/LEPHARE/lephare.html} code \citep{Arnouts99,Ilbert06} by switching on the emission line option. There are ten bands used in the fitting process, including $u^*$ (CFHT), $B_{\rm J}$ (Subaru), $V_{\rm J}$ (Subaru), $g^+$ (Subaru), $r^+$ (Subaru), $i^+$ (Subaru), $i*$ (CFHT), $z^+$ (Subaru), $J$ (UKIRT), and $K$ (CFHT) \citep{Ilbert09}. These ten bands contain PSF-matched photometry data and have similar detection limits. A number of emission lines are involved, i.e. Ly$\alpha$, H$\alpha$, H$\beta$, [OII], and [OIII] lines. The redshift is fixed in this process to improve the SED fitting accuracy, since we have assumed that the given redshifts here are accurate and can be seen as the real redshifts. The templates of extinction curves and other initial parameter setups are  the same as \cite{Ilbert09}. We have checked the fitting results with that provided by the COSMOS catalog, and find that they are in good agreements. The examples of derived SEDs from the ten bands are shown in Figure~\ref{fig:Spec_fit_examples}.

Secondly, we convolve the obtained SEDs with the CSST spectroscopic total transmission curves (in dashed lines shown in the right panel of Figure~\ref{fig:Transmission}), and select the galaxy samples as extended sources with $GV\lesssim22.5$, which is about one magnitude shallower than the 5$\sigma$ point source limit of the $GV$ band. We then obtained about 19,500 galaxies and corresponding SEDs. The mock galaxy redshift distribution of the CSST spectroscopic survey is shown in the left panel of Figure~\ref{fig:spec_photo_zdis}. As can be seen, it has a peak at $z=0.3-0.4$ and can extend to $z=1.5$. We find that both the number density and redshift distribution are consistent with the results derived from the zCOSMOS survey \citep{Lilly07,Lilly09} given in \cite{Gong19}.

Thirdly, we need to consider the instrumental effect and add noise on the SEDs. This requires to analyze and simulate the observational process. Since the CSST spectroscopic observation is performed by slitless gratings, the galaxy light will be dispersed and form multi-order spectra with low wavelength resolution on the detector. Only the first-order spectrum is considered in this work. In addition, galaxy spectra have 2-d feature on the detector, which are along and perpendicular to the dispersion direction. Besides, the point spread function (PSF) also needs to be included to simulate the observed spectrum. Hence, the dispersed galaxy spectral image on the detector can be estimated as \citep{Ubeda11,Cao18}
\be
{\rm SI}(x,y,\lambda)=I(x,y) \otimes {\rm SED_{pix}}(x,y,\lambda) \otimes {\rm PSF}(x,y),
\ee
where $x$ and $y$ denote the position on the detector, and $\lambda$ is the observed wavelength. $I(x,y)$ is the normalized galaxy surface energy distribution function, which can extend and redistribute a galaxy 1-d SED into a 2-d strip based on galaxy spatial energy profile. ${\rm PSF}(x,y)$ is the point spread function, which is assumed to be 2-d Gaussian distribution with the radius of 80\% energy concentration $R_{\rm EE80}\lesssim0.3''$ (about 4 pixels on the detector) \citep{Gong19}. ${\rm SED_{pix}}(x,y,\lambda)$ is the observed detector-pixelized SED with low CSST spectral resolution. It can be derived by a few steps. We first convolve the high-resolution SED obtained from the COSMOS catalog with the CSST spectroscopic total transmissions of the $GU$, $GV$, and $GI$ bands. Then we degrade the ``observed" high-resolution SED to the CSST low-resolution SED. For the CSST spectroscopic survey, a spectral resolution unit $\Delta\lambda$ is designed to averagely take about 4 pixels on the detector for all three spectroscopic bands. The CSST spectral resolution is defined as R=241 at 337.5 nm for $GU$ band,  263 at 525 nm $GV$ band, and 270 at 810 nm $GI$ band, so the $\Delta \lambda$ above are 1.4nm, 2.0 nm and 3.0 nm for the $GU$, $GV$ and $GI$ bands, respectively, by which the SED will be assigned as an 1-d pixel array. So in this way, we can find the relation between pixels (or positions on the detector) and wavelengths, and then ${\rm SED_{pix}}(x,y,\lambda)$ can be obtained.

\begin{figure*}
\centering
\includegraphics[width=3.5in]{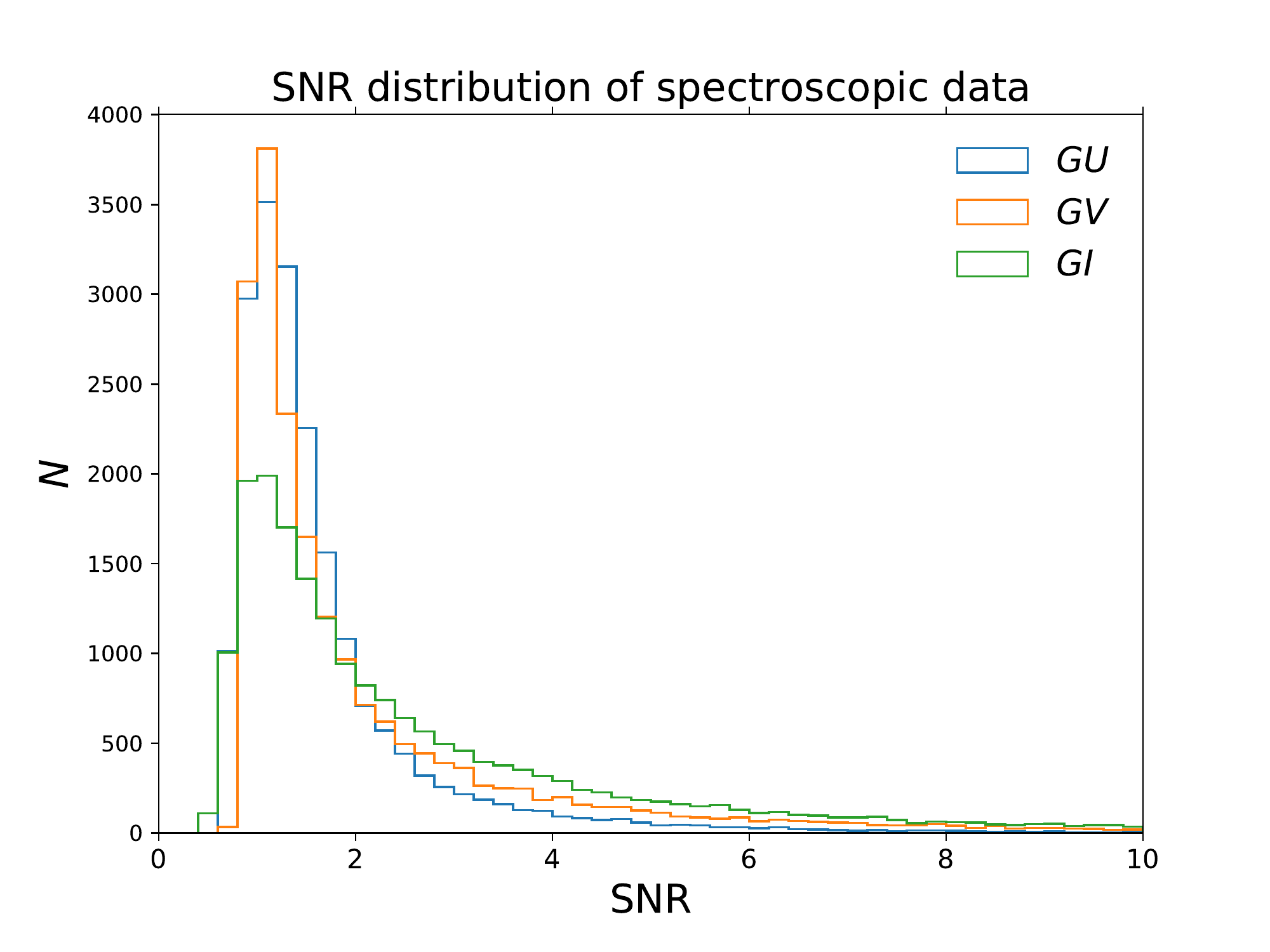}
\includegraphics[width=3.5in]{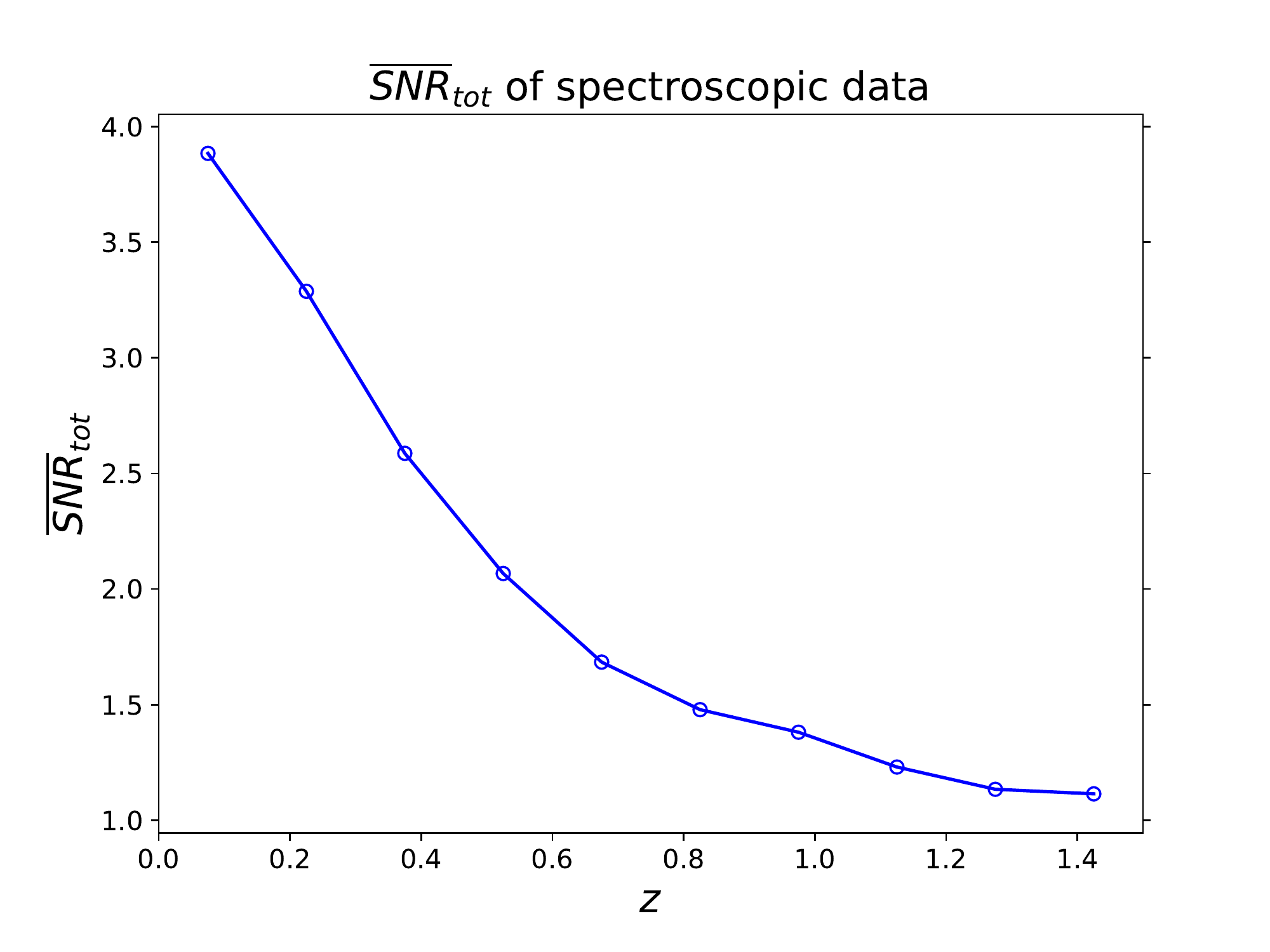}
\includegraphics[width=3.5in]{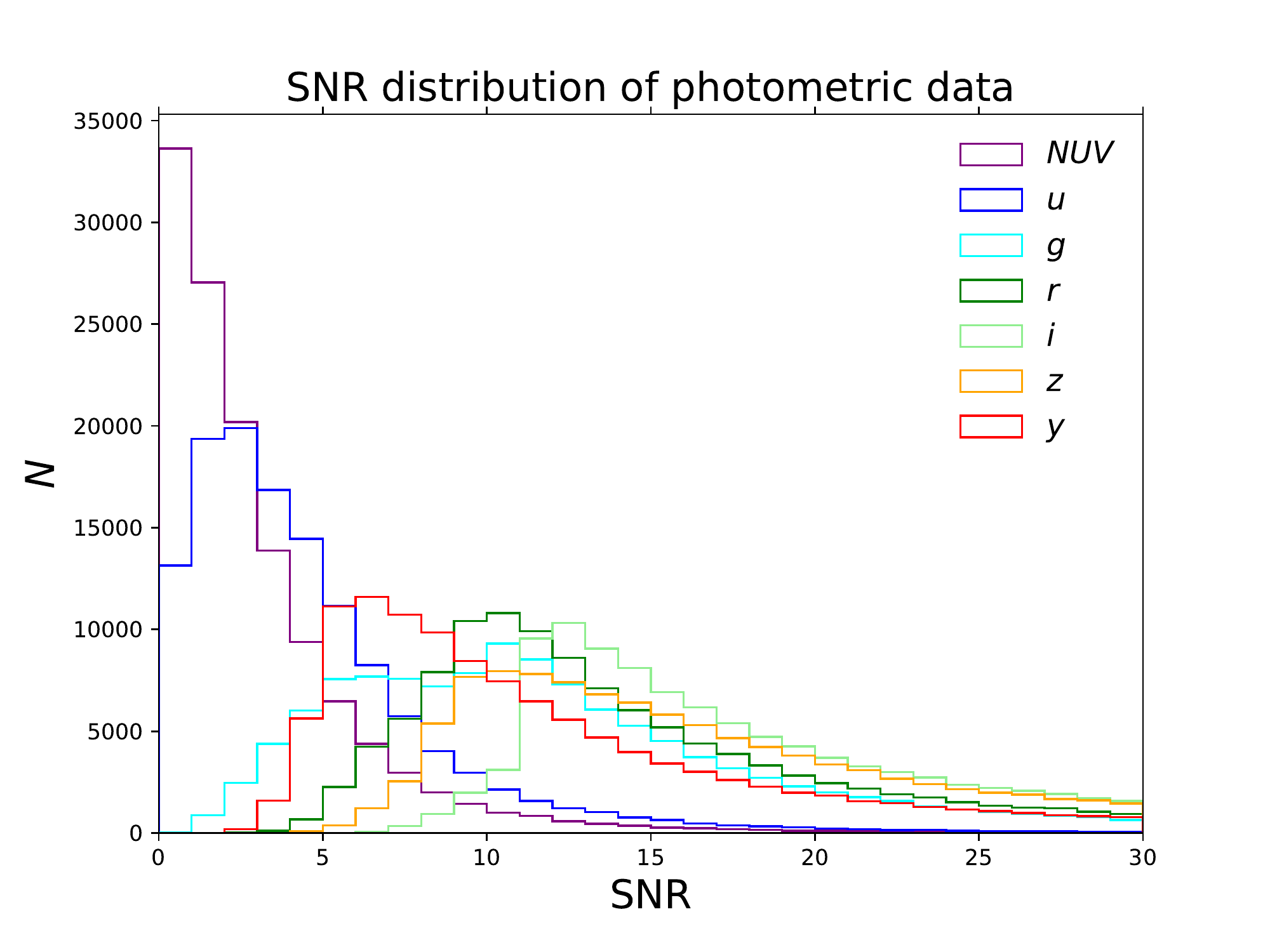}
\includegraphics[width=3.5in]{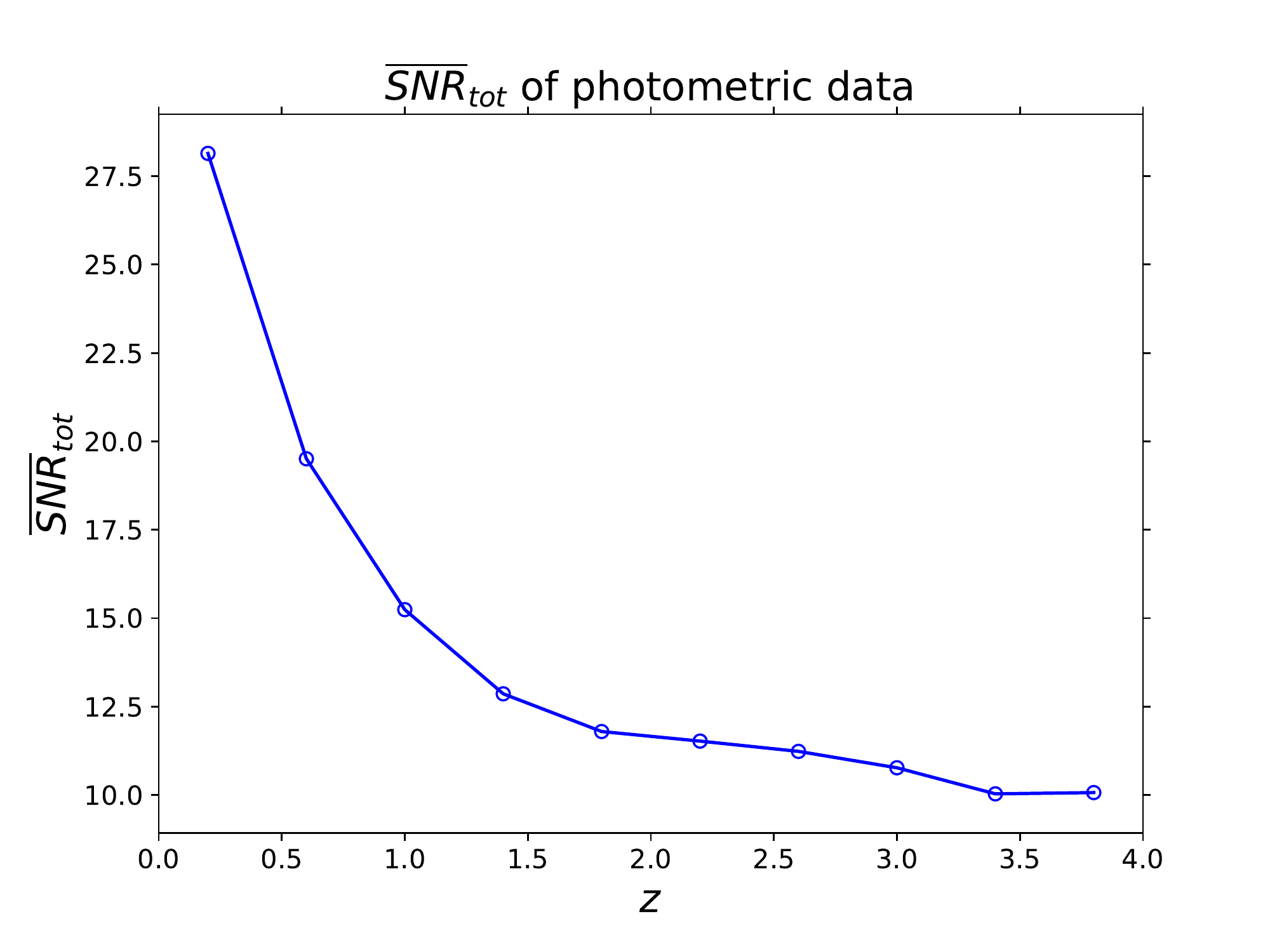}
\caption{{\it Top panels}: the SNR distributions of the mock data for the CSST three spectroscopic bands, and the distribution of the average total SNR of all galaxies for all the three bands at $z$. {\it Bottom panels}: the same cases for the CSST photometric survey.}
\label{fig:spec_photo_snr}
\end{figure*}

\begin{figure*}
\centering
\includegraphics[scale=0.49]{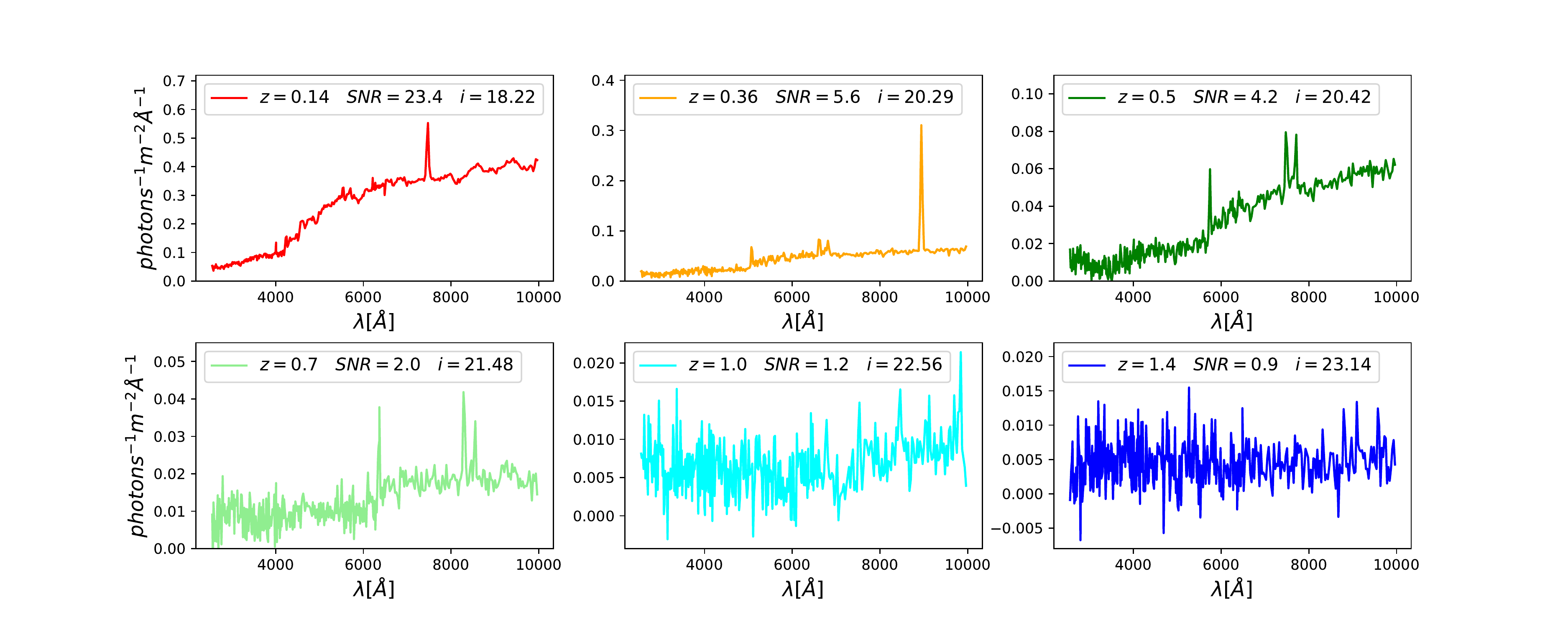}
\caption{Examples of mock slitless spectra measured by the CSST spectroscopic survey at different redshifts. As can be seen, the spectra with low SNRs are dominated by noise, that the emission lines are not obvious and even submerged by noise. In these cases, the SED template-fitting method may not work well, and other methods, like neural network, need to be considered.}
\label{fig:spectrum}
\end{figure*}

\begin{figure}
\centering
\includegraphics[scale=0.45]{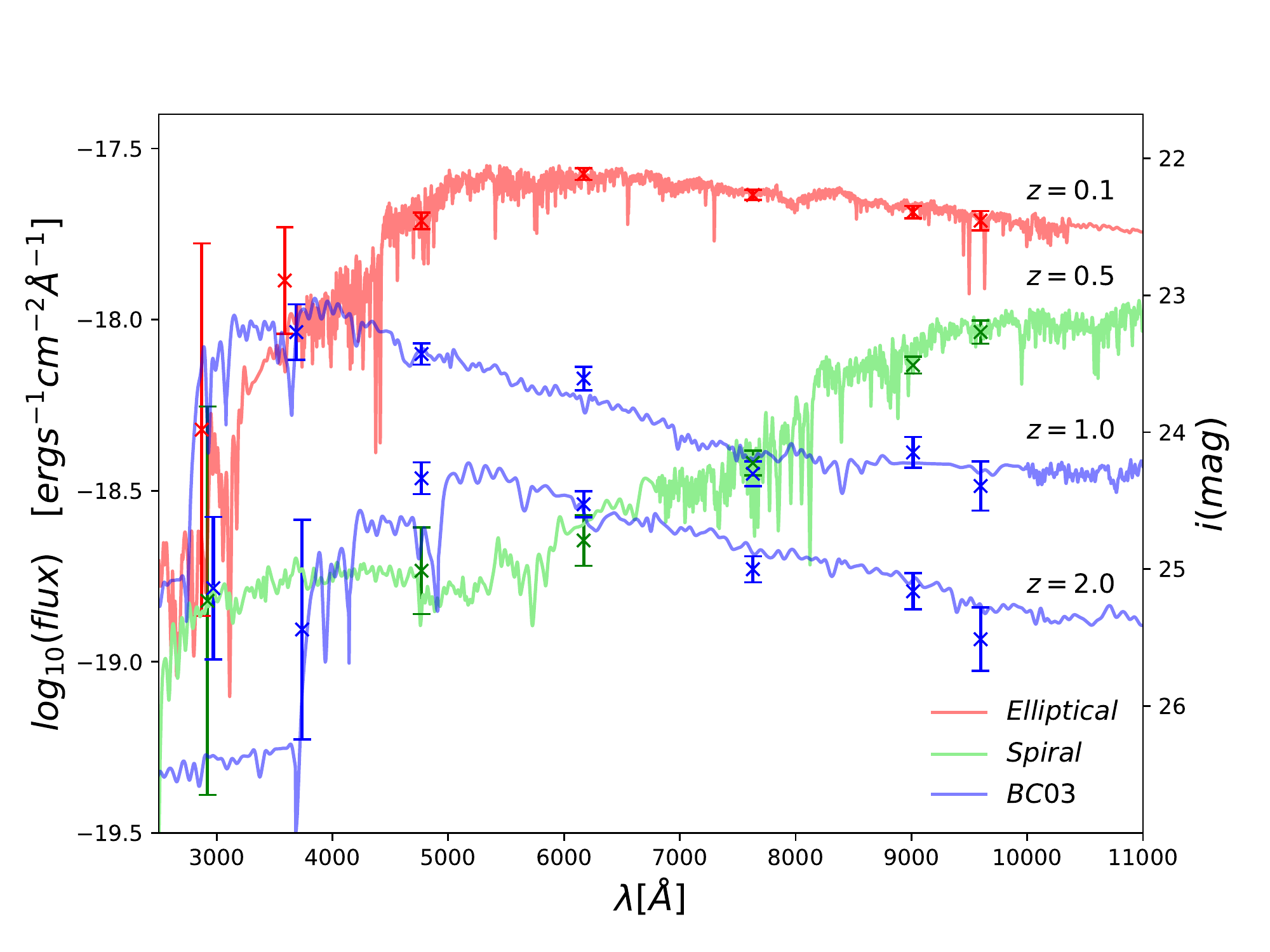}
\caption{Examples of mock flux data at $z=0.1$, 0.5, 1.0, and 2.0 for the seven bands of CSST photometric survey. For comparison, the corresponding SEDs are also shown. The flux data are shifted by a random draw from Gaussian distribution with $\sigma=\sigma_{\rm m}$.}
\label{fig:SEDs}
\end{figure}

After estimating the observed spectral image ${\rm SI}(x,y,\lambda)$, we can evaluate the noise per spectral resolution unit. The SNR can be estimated by \citep{Bohin11,Ubeda11}
\begin{flalign} \label{eq:SNR}
{\rm SNR}&=\frac{C_{\rm s}\,t}{\sqrt{C_{\rm s}t+N_{\rm pix}(B_{\rm sky}+B_{\rm det})t+N_{\rm pix}N_{\rm read}R_{\rm n}^2}}&,
\end{flalign}
where $C_{\rm s}$ is the electron counting rate in ${\rm e}^-{\rm s}^{-1}$, which can be estimated by 
\be \label{eq:C_s}
C_{\rm s}=A_{\rm eff}\int_{\lambda_{\rm min}}^{\lambda_{\rm max}} S_{\rm obs}(\lambda)\tau(\lambda)\frac{\lambda}{hc}d\lambda,
\ee
where $h$ and $c$ are the Planck constant and speed of light, respectively, $A_{\rm eff}$ is the effective telescope aperture area for a band, $S_{\rm obs}(\lambda)$ is the observed SED, and $\tau(\lambda)=m_{\rm eff}^{\rm s}(\lambda)T_{\rm s}(\lambda)$ is the system throughput for the CSST spectroscopic survey. Here $T_{\rm s}(\lambda)$ is the total transmissions of the CSST spectroscopic filters. $m_{\rm eff}^{\rm s}$ is the mirror efficiency, and it is about 0.6 for $GU$, 0.8 for $GV$, and 0.8 for $GI$ band. It is the sum of the signal from all pixels for a spectral resolution unit $\Delta \lambda=\lambda_{\rm max}-\lambda_{\rm min}$. Since the regions of generating emission lines in a galaxy is usually small, for simplicity, we assume that most galaxies can be approximated as point sources in the CSST spectroscopic survey. We can find that this assumption will not affect our results significantly. Then the number of pixels $N_{\rm pix}$ in a spectral resolution unit can be estimated as 4 (along the dispersion direction) $\times$ 8 (perpendicular to the dispersion direction) $=32$ within the $R_{\rm EE80}$\footnote{Based on the instrumental design, the CSST $R_{\rm EE80}$ takes about 4 pixels, so that the diameter takes 8 pixels.}. $t$ is the exposure time given by $150\ {\rm s}\times4=600\ {\rm s}$ \citep{Gong19}. $B_{\rm sky}$ is the sky background which is given by
\be \label{eq:B_sky}
B_{\rm sky}=A_{\rm eff}\int I_{\rm sky}l^2_{\rm p}\tau(\lambda)\frac{\lambda}{hc}d\lambda,
\ee
where $I_{\rm sky}$ is the surface brightness of the sky background in $\rm erg\,s^{-1}cm^{-2}\AA^{-1}$, which mainly includes zodiacal light and earthshine components \citep{Ubeda11}, $l_{\rm p}=0.074''$ is the pixel scale of the detector. We find that $B_{\rm sky}$ is 0.018, 0.212, and 0.288 ${\rm e}^-{\rm s}^{-1}{\rm pixel}^{-1}$ for the $GU$, $GV$, and $GI$ bands, respectively. $B_{\rm det}=0.02\ {\rm e}^-{\rm s}^{-1}{\rm pixel}^{-1}$ is the detector dark current, $N_{\rm read}=4$ is the number of detector readouts, and $R_{\rm n}=5\ {\rm e}^-{\rm pixel}^{-1}$ is the read noise. More details can be found in \cite{Cao18} and \cite{Gong19}. Lastly, the SED noise per resolution unit can be obtained by calculating the denominator of Eq.~(\ref{eq:SNR}).

In Figure~\ref{fig:spec_photo_snr}, we show the SNR distributions of the three CSST spectroscopic bands and average total SNR for all three bands of all galaxies $\overline{\rm SNR}_{\rm tot}$ at $z$ in the top left and right panels, respectively. In the top left panel, we can find that the peaks of SNR distribution for the three bands are all around ${\rm SNR}=1$, and there are few samples at $\rm SNR>6$. In the top right panel, $\overline{\rm SNR}_{\rm tot}$ can be as high as 4 at $z\sim0$, and slowly declines towards to $\sim1$ at $z=1.5$. This indicates that most galaxies have relatively low SNRs in the CSST spectroscopic survey, and it is probably hard to extract their redshifts accurately . This can be more obvious as shown in Figure~\ref{fig:spectrum}, where we show the examples of galaxy slitless spectra for the CSST spectroscopic survey. As can be seen, the noise can be very strong and dominant for the high-$z$ galaxy samples, and the emission lines are easily submerged by the noise when the SNRs are low. This is quite challenging for the ordinary SED template-fitting method to derive accurate redshifts from the data, and other methods (like neural network) are needed to explore here.

Note that the spectra measured by slitless spectroscopic surveys can suffer overlapping problem, that a dispersed multi-order spectrum may overlap with parts of another nearby spectrum, especially in crowded fields. This effect can make trouble in spectrum extraction and suppress the spectral quality and the fraction of usable spectra. However, it is not a main issue in this work, since we assume that the galaxy spectra have already been obtained by previous data processing stages, and no matter overlapping or not, the spectral quality can be reflected by the SNR. A few spectra suffering severe overlapping (e.g. in crowded fields) can be simply discarded in the analysis.

	\begin{figure*}[t]
		\centering
		\includegraphics[scale=0.43]{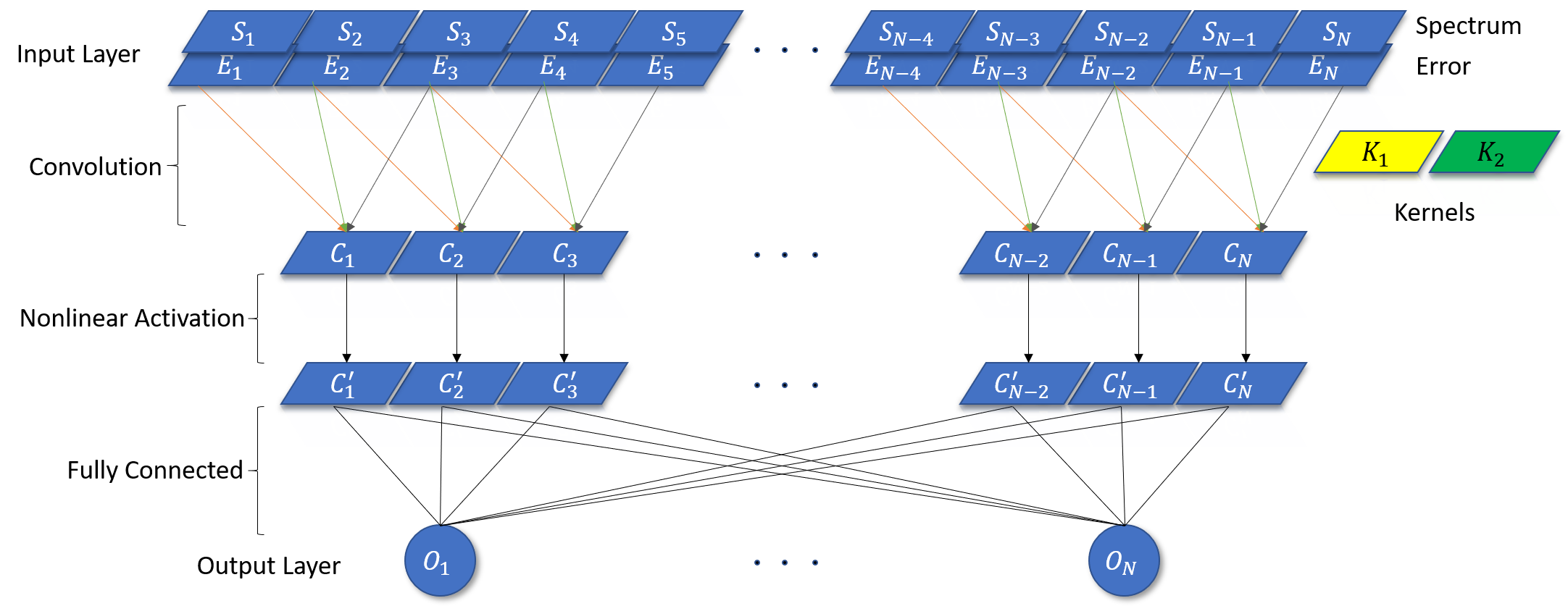}
		\caption{\small{The architecture of 1d-CNN used in the CSST spec-$z$ analysis. The input layer contains two channels, i.e. spectral data and errors. Two convolutional layers are considered, which are operated by two and four kernels, respectively, to obtain feature maps. Two fully connected layers are included to perform the spec-$z$ analysis as a classification task. Output layer contains 750 neurons for classification, which can provide the probability for each redshift interval d$z=0.002$ from $z=0$ to 1.5. Only one convolutional layer and one fully connected layer are shown here as illustration.}}
		\label{fig:CNN_struc}
	\end{figure*}

\subsection{Photometric mock data}
\label{sec:Photometric_data}

The CSST has seven photometric imaging bands covering similar wavelength range as the spectroscopic survey from near-UV to near-IR, i.e. $NUV$, $u$, $g$, $r$, $i$, $z$, and $y$, as shown in solid lines in Figure~\ref{fig:Transmission}. The magnitude 5$\sigma$ limit for point source can be as high as $\sim$26 AB mag for the $g$, $r$, and $i$ bands, and is about $24.5-25.5$ for the other bands. The spatial resolution of the CSST photometric survey is $0.15''$ within 80\% energy concentration for $NUV$, $u$, $g$, $r$ bands, $0.16''$ for $i$ band, and $0.18''$ for $z$ and $y$ bands. More details can be found in  \cite{Cao18} and \cite{Gong19}.

We follow \cite{Cao18} to generate the mock flux data and errors for the seven photometric bands. Firstly, we reproduce the best-fit galaxy SEDs given in the COSMOS catalog, and extend their wavelength range from $\sim900$ to $\sim90\, \rm\AA$ to match the CSS-OS large wavelength coverage \citep{Cao18}. Secondly, we add the dust extinction from interstellar dust in galaxies and absorption of intergalactic medium (IGM) to the galaxy SEDs we obtained. Five extinction laws are considered for the interstellar dust origined, including the ones derived from the Milky Way \citep{Allen76,Seaton79}, Large Magellanic Cloud \citep{Fitzpatrick86}, Small Magellanic Cloud \citep{Prevot84,Bouchet85}, and starburst galaxy \citep{Calzetti00}. For the IGM absorption, we use the attenuation laws computed by \cite{Madau95}. Thirdly, we rescale the obtained SEDs based on the Subaru $i^+$ band magnitude, and convolve them with the CSST filter transmission curves (see the solid lines in the right panel of Figure~\ref{fig:Transmission}) to get the mock CSST photometric flux data. Note that, similar as the CSST spectroscopic survey, the SED here is the observed SED without instrumental effect, and can be seen as high-quality spectral data obtained by powerful spectroscopic surveys.

Next, we can estimate the flux errors for the seven bands. The same SNR formula shown in Eq.~(\ref{eq:SNR}) is used in the calculation. Here the electron count rate can be estimated by Eq.~(\ref{eq:C_s}), but replace $\tau(\lambda)$ by $\tau_{\rm p}(\lambda)=m_{\rm eff}^{\rm p}(\lambda)T_{\rm p}(\lambda)$, which is the system throughput for the CSST photometric survey. $T_{\rm p}(\lambda)$ is the total transmissions of the CSST photometric filters. $m_{\rm eff}^{\rm p}$ is the mirror efficiency, and it is about 0.5 for $NUV$, 0.7 for $u$, and 0.8 for the other bands. $\lambda_{\rm min}$ and $\lambda_{\rm max}$ cover the whole band range. The exposure time $t=150\ \rm s\times2=300$ s, the number of detector pixels covered by an object $N_{\rm pix}$ can be derived by galaxy size, the number of detector readout $N_{\rm read}=2$, and the values of $B_{\rm det}$ and $R_{\rm n}$ are the same as the CSST spectroscopic survey. The sky background $B_{\rm sky}$ here can be estimated by Eq.~(\ref{eq:B_sky}) using $\tau_{\rm p}(\lambda)$. We find that $B_{\rm sky}$ are 0.003, 0.018, 0.156, 0.200, 0.207, 0.123, and 0.036 $\rm e^-s^{-1}pixel^{-1}$ for $NUV$, $u$, $g$, $r$, $i$, $z$, and $y$ bands, respectively. Then the photometric error can be calculated as $\sigma_{\rm p}\simeq\rm 2.5 log_{10}(1+1/SNR)$ mag \citep{Bolzonella00,Pozzetti96,Pozzetti98}. We assume the systematic error $\sigma_{\rm sys}=0.02$, which can be seen as a systematic floor for the photometric surveys \citep[e.g. see][]{Astier06, Sullivan06}. The total magnitude error is given by $\sigma_{\rm m}=\sqrt{\sigma_{\rm ph}^2+\sigma_{\rm sys}^2}$ \citep{Cao18}.

Since accurate photo-$z$ is needed in the weak lensing and 2-d galaxy clustering analysis, only high-quality photometric data are selected in real observations. Here we select the flux data with $\rm SNR\ge10$ for $g$ or $i$ band, and lastly about 126,000 galaxies are obtained \citep{Cao18}. In the right panel of Figure~\ref{fig:spec_photo_zdis}, the redshift distribution of the selected mock galaxy sample for the CSST photometric survey has been shown, which has a peak at $z=0.7$ and can extend to $z=4$. In the bottom panels of Figure~\ref{fig:spec_photo_snr}, we show the SNR distributions for the seven photometric bands, and the average total SNR (i.e. $\rm \overline{SNR}_{tot}$) for all bands and all galaxies at $z$. In the bottom left panel, we can find that the peaks of SNR distributions are less than 5 for $NUV$ and $u$ bands, $\sim7$ for $y$ band, and $\sim10$-13 for $g$, $r$, $i$, and $z$ bands. In the bottom right panel, $\rm \overline{SNR}_{tot}$ can be greater than 25 at $z<0.5$ and declines to about 10 (the SNR cut for $g$ or $i$ band) at $z=4$. In Figure~\ref{fig:SEDs}, we show examples of mock flux data for the seven CSST photometric bands at $z=0.1$, 0.5, 1.0, and 2.0. The corresponding observed SEDs are also shown for comparison. We can see that,  except for $NUV$ and $u$ bands, the flux data have relatively small errors, and they can well represent the features of galaxy SEDs.

\section{Neural Network}\label{sec:Neural Network}

After obtaining the mock data of the CSST spectroscopic and photometric surveys, we can explore the estimation and accuracy of the spec-$z$ and photo-$z$ using the neural network. The CNN and MLP are adopted in the spec-$z$ and photo-$z$ analysis, respectively, and the details of their architectures and training processes are discussed in this section.

\subsection{Network architecture}

	\renewcommand{\arraystretch}{1.5}
	\begin{table}
		\caption{Details of 1d-CNN used in the CSST spec-$z$ analysis.}
		\label{tab:CNN_para}
		\begin{center}
		\begin{tabular}{lll}
			\hline
			Layers & Output status$^a$ & Number of paras.$^b$\\
			\hline
			\hline
			Input & (369,2) & 0\\
			\hline
			Conv1D (2,5)$^c$\ \ \ \ \ \ \   & (369,2) & 22\\
			\hline
			ReLU & (369,2) & 0 \\
			\hline
			MaxPooling & (184,2) & 0 \\
			\hline
			Conv1D (4,5)$^d$ & (180,4) & 44 \\
			\hline
			ReLU & (180,4) & 0 \\
			\hline
			MaxPooling & (90,4) & 0 \\
			\hline
			Flatten & (360,1) & 0 \\
			\hline
			FC$^e$ & (256,1) & 92,416 \\
			\hline
			ReLU & (256,1) & 0 \\
			\hline
			FC$^e$ & (256,1) & 65,782 \\
			\hline
			ReLU & (256,1) & 0\\
			\hline
			Output & (750,1) & 192,750\\
			\hline
			\hline
		\end{tabular}
		\end{center}
		\vspace{-2mm}
		$^a$ The format is (number of data points or neurons, number of channels or dimensions).\\
		$^b$ Total number of parameters: 351,024.\\
		$^c$ Two convolution kernels with spatial dimension $1\times5$.\\
		$^d$ Four convolution kernels with spatial dimension $1\times5$.\\
		$^e$ FC: fully connected layer. \\
	\end{table}

The CNN and MLP adopted in this work are implemented by Keras\footnote{https://keras.io}, a high-level neural network application programming interface (API) running on top of TensorFlow\footnote{https://www.tensorflow.org}. We use 1-d CNN to perform spec-$z$ analysis for the CSST slitless spectroscopic survey. The CNN has advantages of capturing local information and reducing the model complexity and overfitting problem, so it is suitable and widely used in extraction of feature information from images. Since galaxy spectra can be seen as 1-d images,  we apply the 1-d CNN in the spec-$z$ estimation.

In Figure~\ref{fig:CNN_struc}, the architecture of the 1-d CNN adopted in this work has been shown, which is mainly composed of input layer, convolutional layers, fully connected layers, and output layer. In the input layer, we make use of two channels, i.e. spectral data and errors. As we can see in the later discussion, the errors as input data can significantly improve the accuracy of the spec-$z$. We totally have 369 data points for each galaxy spectrum, which has wavelength intervals as 14, 20, and 30 $\rm\AA$ for the $GU$, $GV$, and $GI$ bands, respectively. 

We set two 1-d convolutional layers, and this is found to be good enough to derive the results (i.e. more layers get similar results). For the first and second convolutional layers, two and four kernels are employed to extract useful features of the data, respectively. All kernels have the same $1\times5$ spatial dimension. To avoid losing marginal information, we set the same padding on the first convolutional layer to obtain a feature map with the same number of elements as the input data. After each convolutional layer, the Rectified Linear Unit (ReLU) is applied as nonlinear activation function \citep{Nair10}, which is given by $y={\rm max}(0,x)$, to implement non-linearity, reduce vanishing-gradient problem, and speed up the training process. The pooling layer is followed to downsample the feature map. The max pooling method is adopted here, that the maximum value is chosen in two adjacent elements. Then we obtain a feature map with a dimension of 90$\times$4. Next flattening is applied to get a 1-d vector for connecting with the fully connected layer by trainable weights.

	\begin{figure}
		\centering
		\includegraphics[scale=0.45]{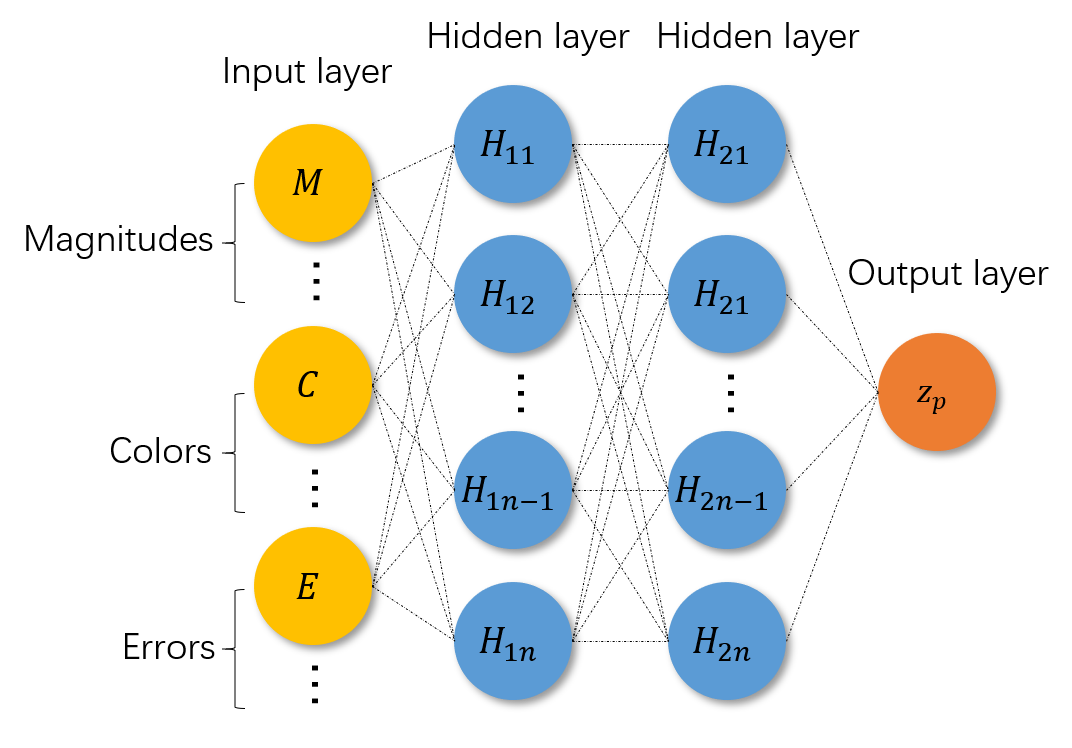}
		\caption{The architecture of the MLP used in the CSST photo-$z$ estimation. The values included in the input layer are 7 magnitudes and errors, and 6 colors. Two hidden layers are structured to predict photo-$z$ given by the output layer.}
		\label{fig:ANN_struc}
	\end{figure}

	\renewcommand{\arraystretch}{1.5}
	\begin{table}
		\caption{Details of the MLP used in the CSST photo-$z$ analysis.}
		\label{tab:ANN_para}
		\begin{center}
		\begin{tabular}{lll}
			\hline
			Layers \ \ \ \ \ \ \ \ \ \ \ \ \ \  \ & Output status$^a$ & Number of paras.$^b$\\
			\hline
			\hline
			Input & 20 & 0 \\
			\hline
			FC$^c$ & 40 & 840\\
			\hline
			ReLU & 40 & 0\\
			\hline
			FC$^c$ & 40 & 1640\\
			\hline
			ReLU & 40 & 0\\
			\hline
			Output & 1 & 41\\
			\hline
			\hline
		\end{tabular}
		\end{center}
		\vspace{-2mm}
		$^a$ Number of data points or neurons.\\
		$^b$ Total number of parameters: 2,521.\\
		$^c$ FC: fully connected layer.\\
	\end{table}

We structure two fully connected layers, and the ReLU is applied after each. In the training process, in order to reduce overfitting, a dropout layer is added with a keeping rate of 0.5 after each fully connected layer \citep{Srivastava14}, which means half of neurons are temporarily turned off in training. As we discuss later, the dropout can affect the accuracy in the training and validation process. The fully connected layer also can be seen as a classifier, which can deal with the spec-$z$ analysis as a classification task \citep{Stivaktakis19}. In details, the CSST spectroscopic redshift range from $z$=0 to 1.5 is divided by 750 bins or classes with redshift interval d$z$=0.002\footnote{We notice that, as shown in the left panel of Figure~\ref{fig:spec_photo_zdis}, in some of these narrow redshift bins at $z\gtrsim1.3$, the number of galaxies may be close to zero, which can affect training process and hence the spec-$z$ estimate in these bins. However, since there are not many spectroscopic sample in this range for the CSST, and the data from other spectroscopic surveys can be used as training sample, this problem is solvable.}. Correspondingly, in the output layer, 750 neurons are assigned with the multi-class activation function, i.e. softmax function, which takes the exponential form $S_i={\rm exp}(V_i)/\sum_j {\rm exp}(V_j)$. Here $V_i$ is the value of $i$th neurons of output layer. The softmax function can give probability for each class, and the sum is equal to 1. The details of layers, e.g. number of data points, neurons and parameters, can be found in Table~\ref{tab:CNN_para}.
	
\begin{figure*}
\centering
\includegraphics[width=3.5in]{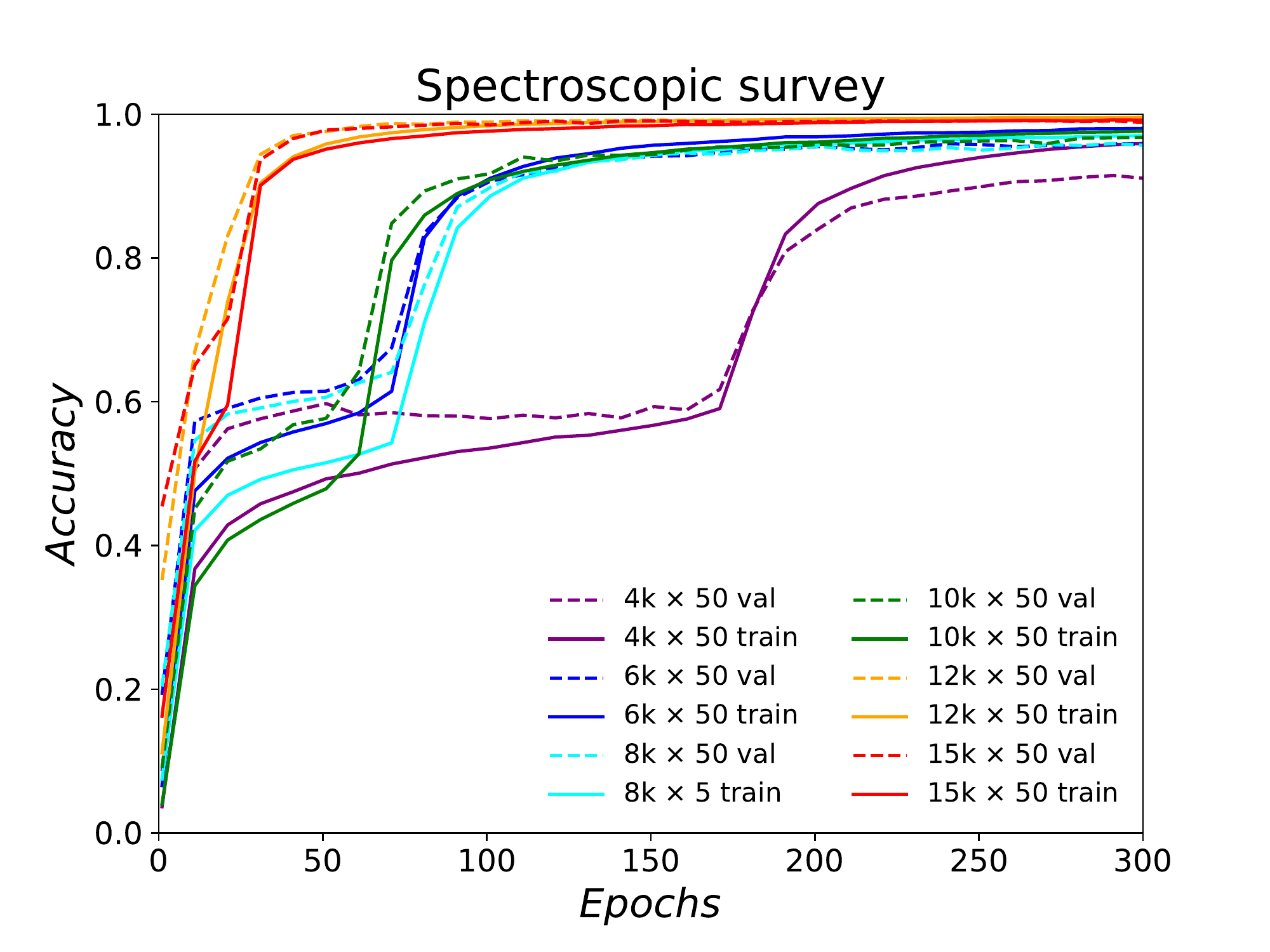}
\includegraphics[width=3.5in]{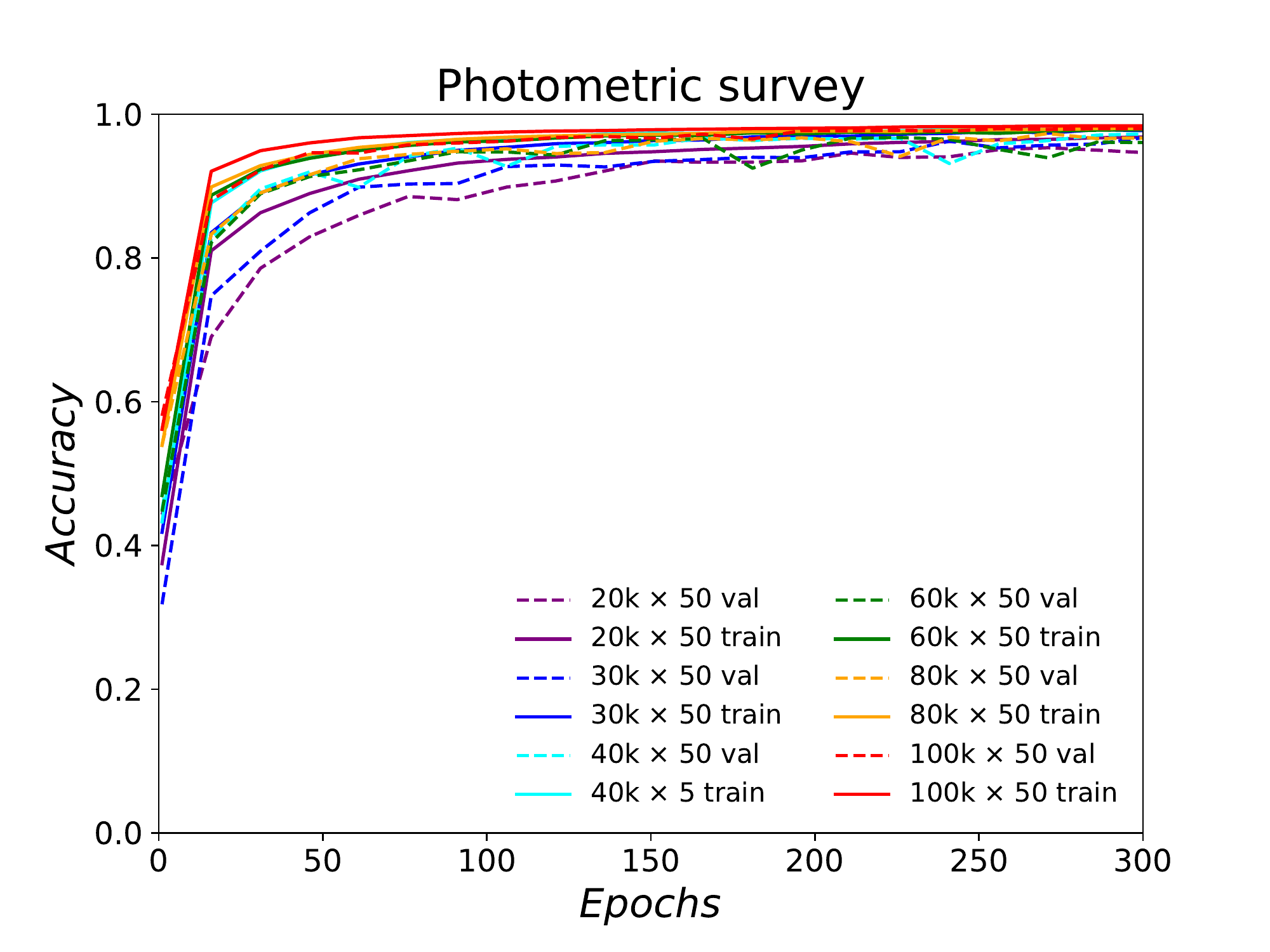}
\caption{\emph{Left}: The fraction of galaxies that can reach the accuracy criteria for the training and validation samples as a function of training epochs in the CSST spectroscopic survey. The results of different number of data from 4,000$\times$50 to 15,000$\times$50 have been shown for the training (solid) and validation (dashed) datasets. 50 realizations are generated from a Gaussian distribution based on the original datasets. \emph{Right}: The results for the photometric survey, and the results for the numbers of training data are from 20,000$\times$50 to 100,000$\times$50 have been shown.}
\label{fig:spec_photo_acc}
\end{figure*}

The MLP is adopted to derive photo-$z$ from the CSST photometric survey, since the CSST photometric data only has seven data points for each galaxy, and is much simpler than the spectroscopic data. We show the MLP architecture and details in Figure~\ref{fig:ANN_struc} and Table~\ref{tab:ANN_para}. In the input layer, we consider the magnitudes, errors, and colors for the seven CSST photometric bands, which gives 20 data points for each galaxy. Two hidden layers are used here. Since we apply the classic structure with $n:2n:2n:1$ feature, where $n$ is the number of data elements, there are 40 neurons in each hidden layer. The ReLU is also applied after each hidden layer. Then the photo-$z$ will be obtained in the output layer. 

\subsection{Training process}

As we discussed in the last section, majority of the CSST slitless spectral data may have relatively large noises and low SNRs (i.e. $\rm SNR<3$), especially for the galaxies at high redshifts (see the top panels of Figure~\ref{fig:spec_photo_snr}). This means that there will be large deviation of the spectral data points from the real spectral curve, that multi-measurements may result in distinctly different shapes for the same spectrum or spectral type due to statistical reason. This problem can be serious for the ordinary SED template fitting method, which may lead to different spec-$z$ predications for the same galaxy with large errors. On the other hand, since all of these measured spectral data with different shapes have statistical origins, we can generate mock data to simulate this feature, and input them into the neural networks as training sample to reduce this effect. Looking from another aspect, these realizations also can be seen as spectra measured from different galaxies with similar spectral type at the same redshift, that can dramatically expand the training sample.

Since the spectral data with high SNRs and resolutions are needed to generate the training data for the CSST slitless spectroscopic survey, it could be still hard to achieve the requirements even for the future powerful spectroscopic surveys at high redshifts $z>1$. Therefore, here we make a conservative assumption that the redshift distribution of the high-quality spectral data used to generate the training data is similar as the CSST spectroscopic survey. This means we assume there are not many data can be used to train the network at $z>1$, and we will generate the training data directly from our mock CSST spectral catalog. If more high-quality spectral data at high-$z$ can be used for training, the results will be undoubtably improved.

We divide the mock spectral data into two parts, i.e. training and testing datasets. We randomly select 15,000 galaxies from the mock spectral dataset as training sample, and the rest of $\sim4500$ are for testing. Assuming that the deviation of the spectral data point at each wavelength follows Gaussian distribution with the spectral error as the $1\sigma$, we randomly generate 50 realizations for each spectrum of the training set. We have checked that more realizations would not improve the result significantly, and the results would not considerably change if the number of realizations is greater than 30. In this training process, the categorical cross-entropy loss function is optimized by Adam optimizer \citep{Kingma14}. It is an efficient method for stochastic optimization that only requires first-order gradients, and can adjust the learning rate automatically. This makes our code more efficient than the current machine learning codes used in the redshift estimation, such as ANNz \citep{Collister04}, ArborZ \citep{Gerdes10}, and CuBANz \citep{Samui17}, in the training process. Based on the pre-running tests, we set the maximum epochs of the neural network to be 300 as the termination criteria. Our network can save the best model of these epochs with highest accuracy during training, which will be applied to the testing sample.

To show the training process straightly and clearly, we define an accuracy criteria as $|z_{\rm pred}-z_{\rm true}|/(1+z_{\rm true})\le 0.002$, where $z_{\rm pred}$ and $z_{\rm true}$ are the predicted and true redshifts, respectively. In the left panel of Figure~\ref{fig:spec_photo_acc}, we show the fraction of the training and validation sample that can achieve the accuracy criteria as a function of training epochs for different sizes of training data. The validation sample used here is set to be the same as the testing sample, which contains 50 Gaussian random realizations of $\sim$4,500 original galaxy spectra. Note that the number of realizations of the validation data is not necessarily the same as the training data. No matter how many realizations of the validation data we created, the results would not change as long as the realizations of training data are more than 30$\sim$50. The results for training and validation data are shown in solid and dashed curves, respectively. We find that the accuracy fractions of all training datasets we check can approach to 1 at epoch$\simeq$300, which means that the spec-$z$ can be efficiently derived in high accuracy by our network. More training data can speed-up this process, and it converges when the number of training data reaches $\sim$12000$\times$50. On the other hand, less training data will take more epochs, and may experience ``metastable state'' but can reach higher accuracy in later epochs. 

Besides, we also notice that the accuracy of the validation data can be higher than that of the training data in early epochs. This is mainly due to two reasons. The first one is the main reason, that the dropout layers are only applied in the training stage, and only half of neurons are used to derive spec-$z$ when training, while all neurons are used in the validation. Hence, in principle, the network is more powerful in the validation and can obtain higher accuracy at the same epoch. The second reason is caused by the difference when calculating the training and validation accuracy. The training accuracy is estimated by averaging the accuracies of all batches in an epoch. Generally speaking, the training accuracy over the first batches are lower than over the last batches. On the other hand, the validation accuracy is obtained using the model generated at the end of epoch, resulting in higher accuracy. In the early epochs, these effects are more obvious since the network is not well trained, while they can be effectively relieved in the late epochs.

For the CSST photometric survey, the spectral data at high redshifts $z\sim4$ could be found in the future high-resolution spectroscopic surveys. As similar as the spectroscopic survey, we assume the redshift distribution of the high-quality spectral data used to generate the photo-$z$ training data is similar as the CSST photometric survey, and can be obtained from the CSST mock photometric catalog. We divide the mock dataset of the CSST photometric survey into the training and testing samples. We randomly selected 10,000 galaxy spectra as testing data, and the rest $\sim$116,000 are employed as training data. Then 50 Gaussian random realizations are created for each galaxy using the same method as the spectroscopic survey, and more realizations would not change the results significantly. The accuracy criteria defined in the photometric survey is given by $|z_{\rm pred}-z_{\rm true}|/(1+z_{\rm true})\le 0.05$. In the right panel of Figure~\ref{fig:spec_photo_acc}, we show the accuracy of training and validation samples versus the training epochs for different number of training data. We can find that the accuracies of all datasets can approach to unity at early epochs $\lesssim100$, which means the network is efficient to find the correct photo-$z$ in high accuracy. Note that the problem of training accuracy less than validation accuracy does not appear here, since no dropout layer is structured in the MLP.

	\begin{figure}[t]
		\centering
		\includegraphics[scale=0.31]{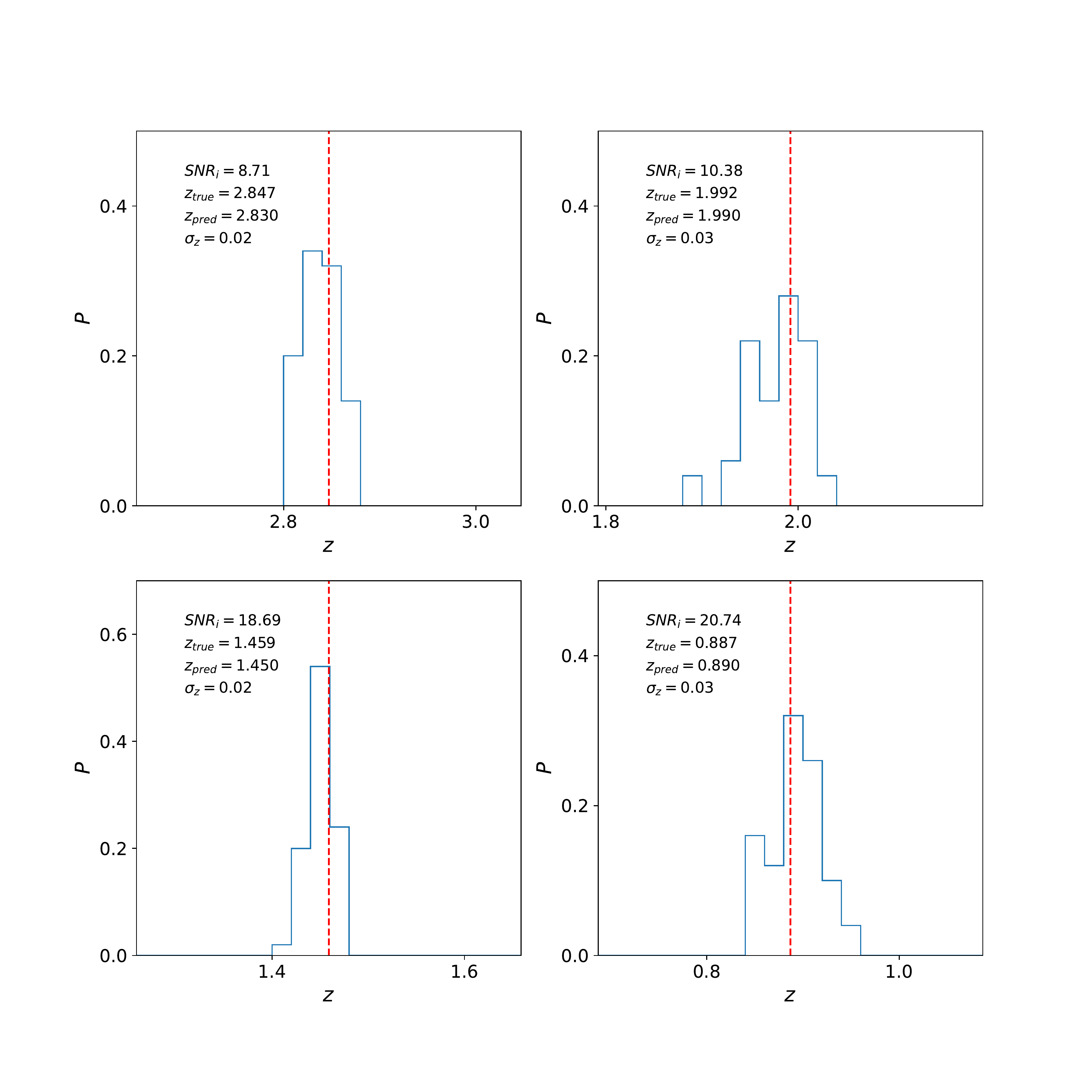}
		\caption{Examples of the photo-$z$ PDFs derived from our neural networks for different $i$-band SNRs. The true photo-$z$ is denoted by the vertical red dashed line. We can find that the width of the PDF is not sensitive to the SNR.}
		\label{fig:photo_z_pdf}
	\end{figure}

	\begin{figure*}[ht]
		\centering
		\subfigure{
			\begin{minipage}[t]{0.31\linewidth}
				\centering
				\includegraphics[scale=0.31]{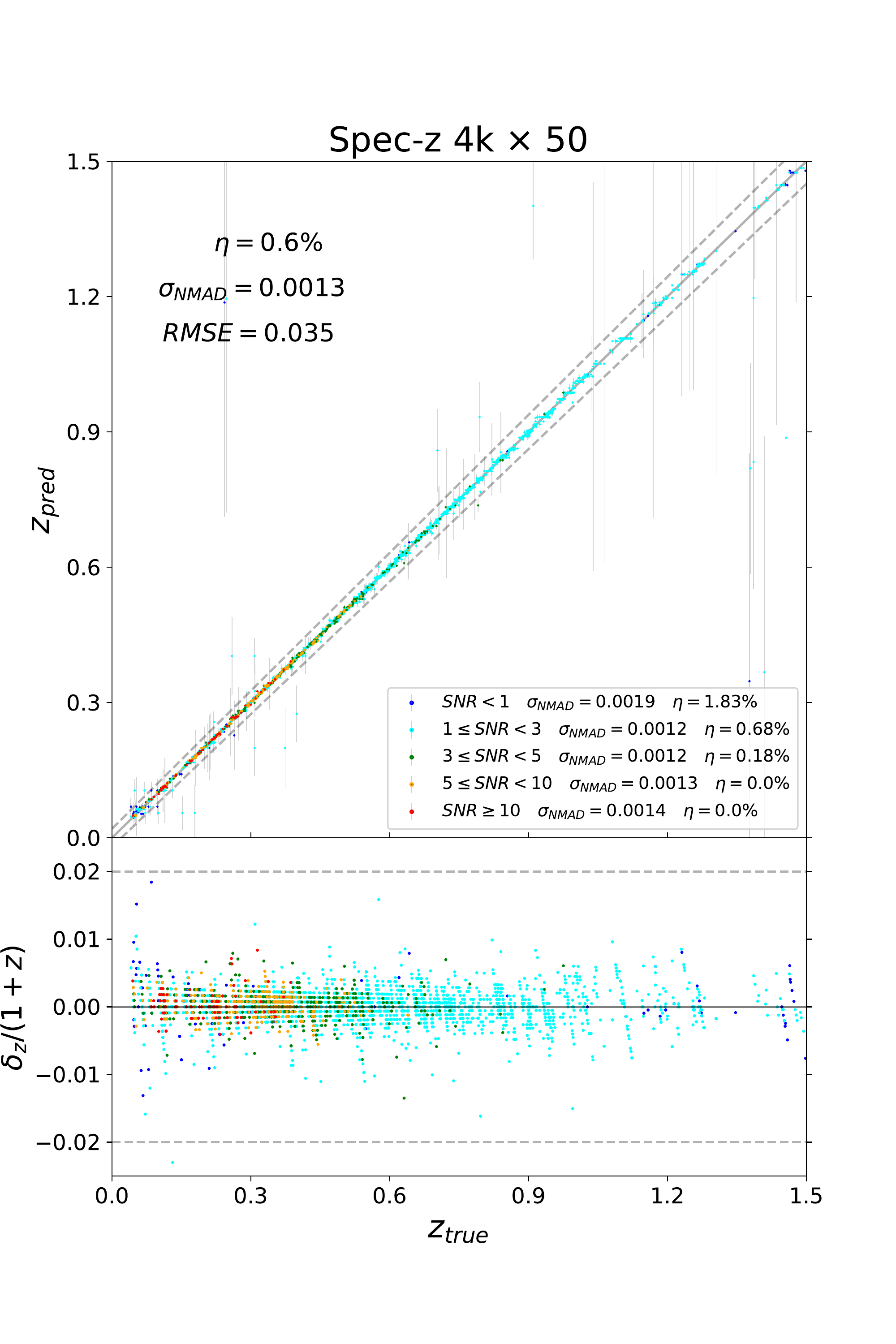}
			\end{minipage}
		}
		\subfigure{
			\begin{minipage}[t]{0.31\linewidth}
				\centering
				\includegraphics[scale=0.31]{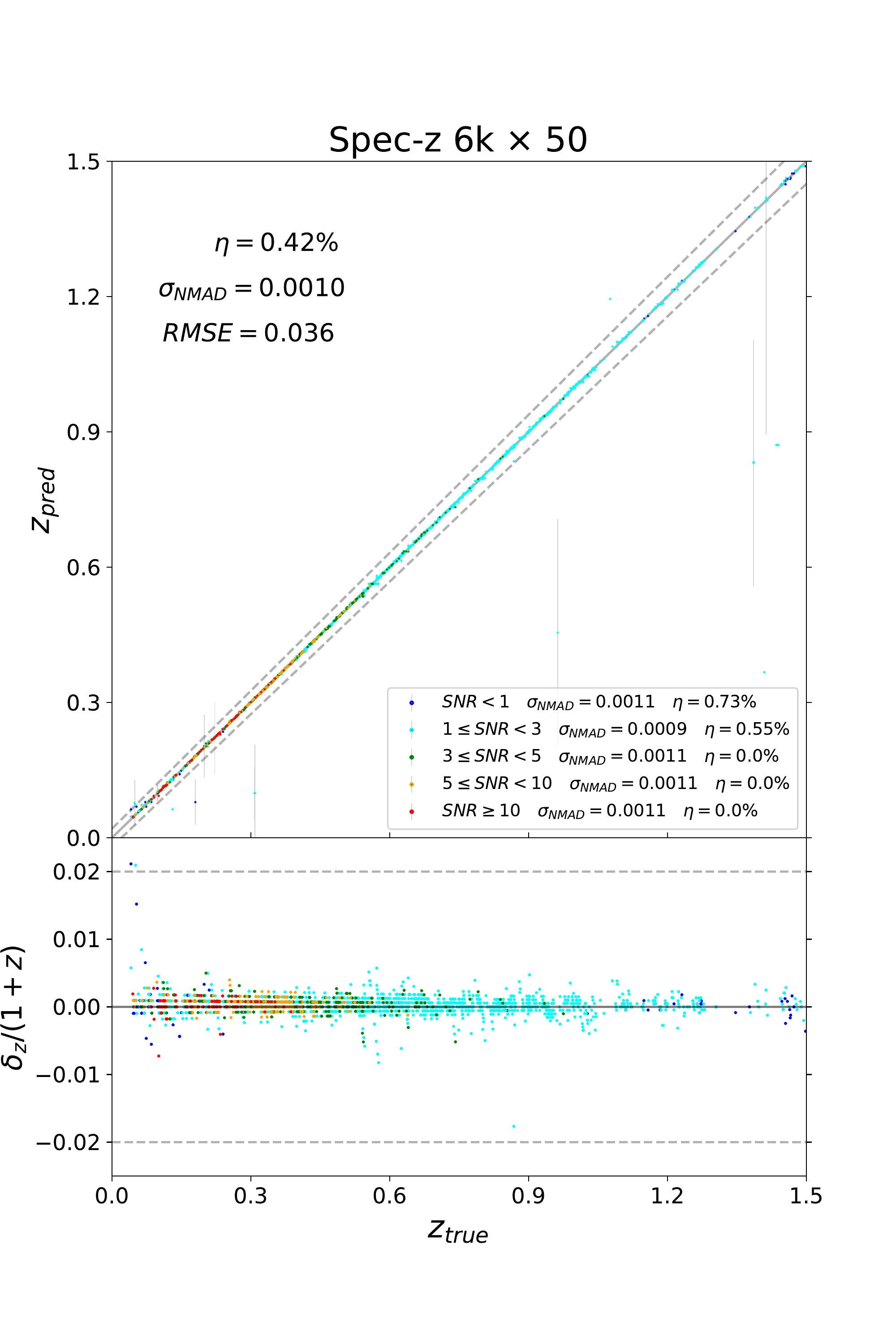}
			\end{minipage}
		}
		\subfigure{
			\begin{minipage}[t]{0.31\linewidth}
				\centering
				\includegraphics[scale=0.31]{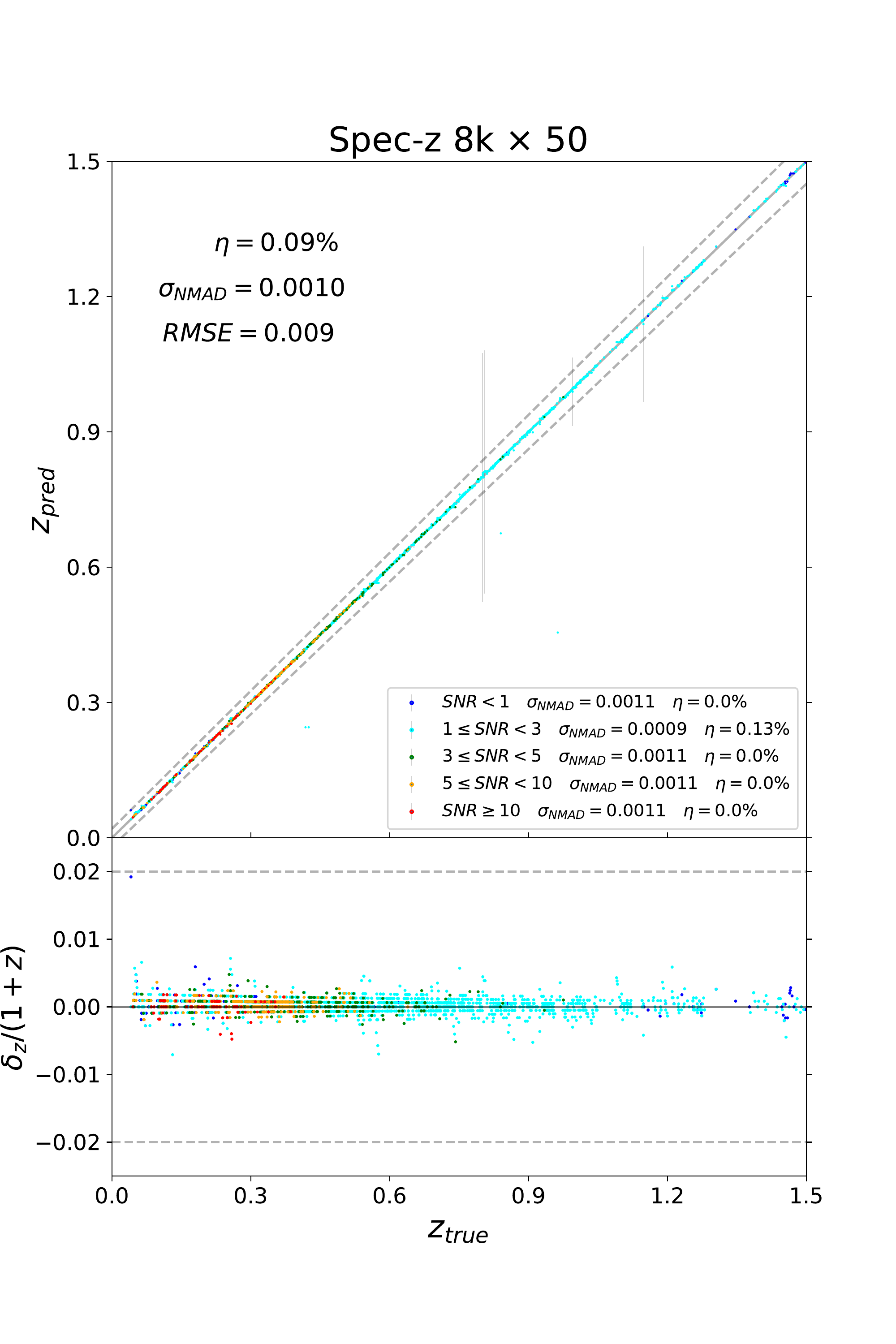}
			\end{minipage}
		}
		
		\subfigure{
			\begin{minipage}[t]{0.31\linewidth}
				\centering
				\includegraphics[scale=0.31]{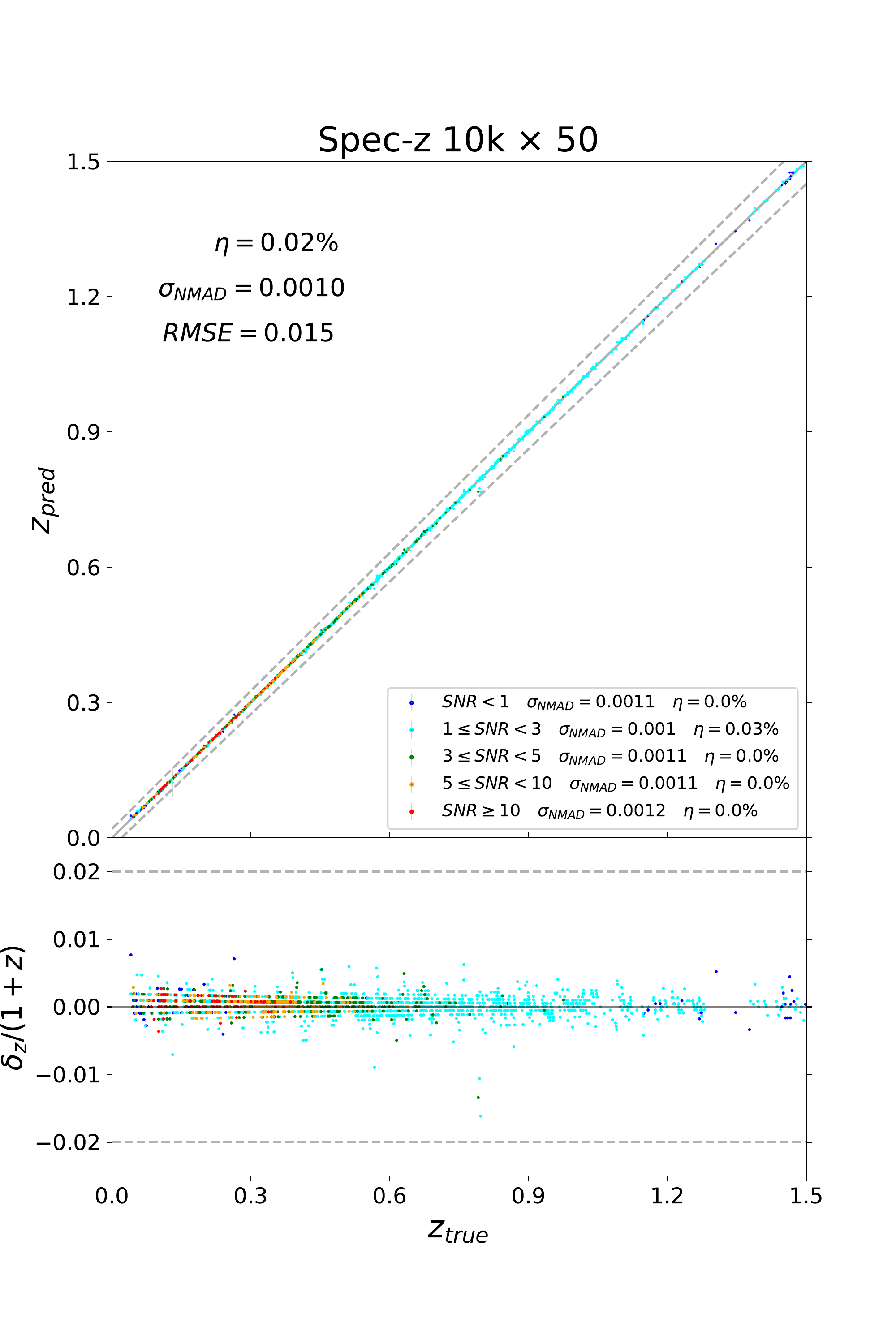}
			\end{minipage}
		}
		\subfigure{
			\begin{minipage}[t]{0.31\linewidth}
				\centering
				\includegraphics[scale=0.31]{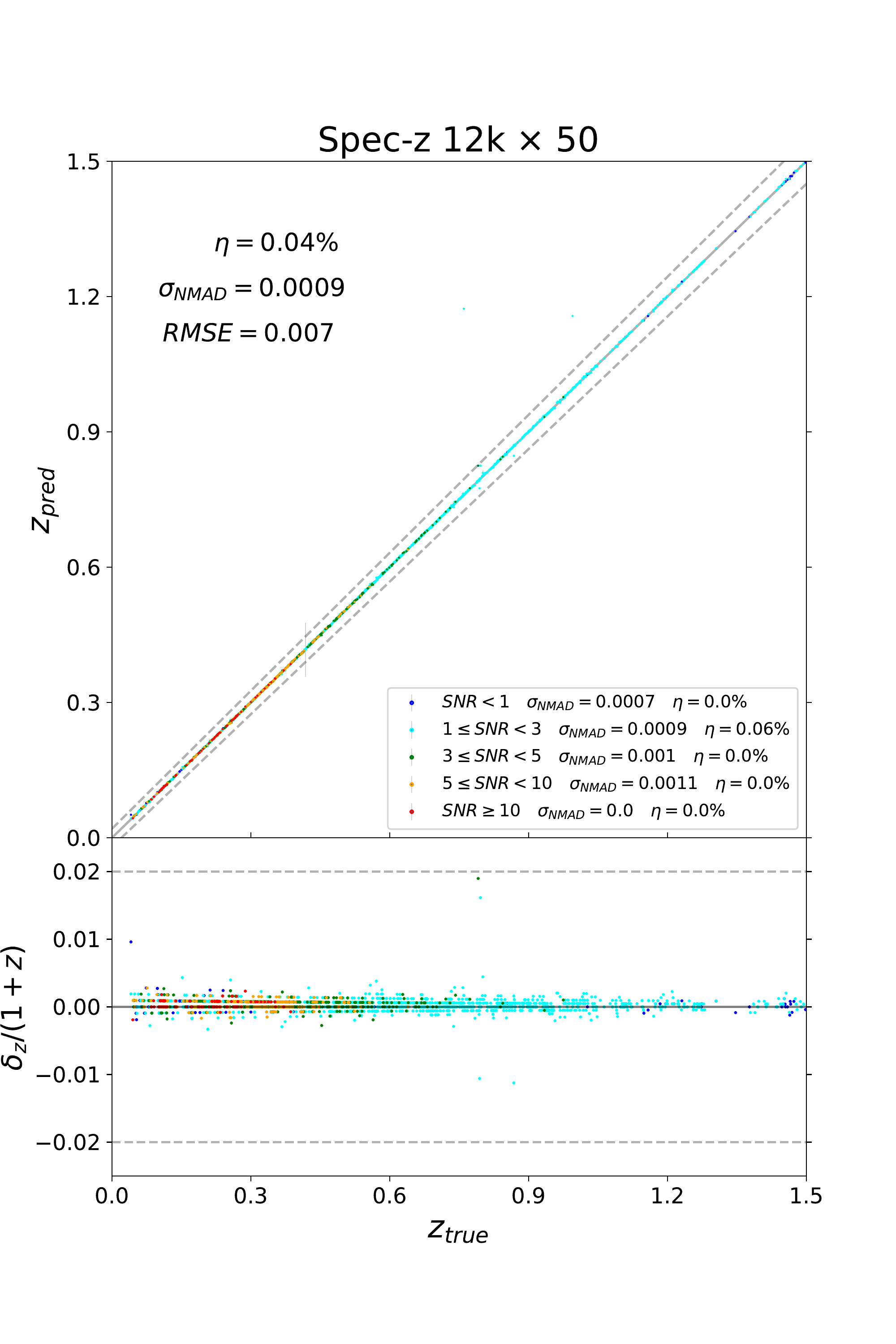}
			\end{minipage}
		}
		\subfigure{
			\begin{minipage}[t]{0.31\linewidth}
				\centering
				\includegraphics[scale=0.31]{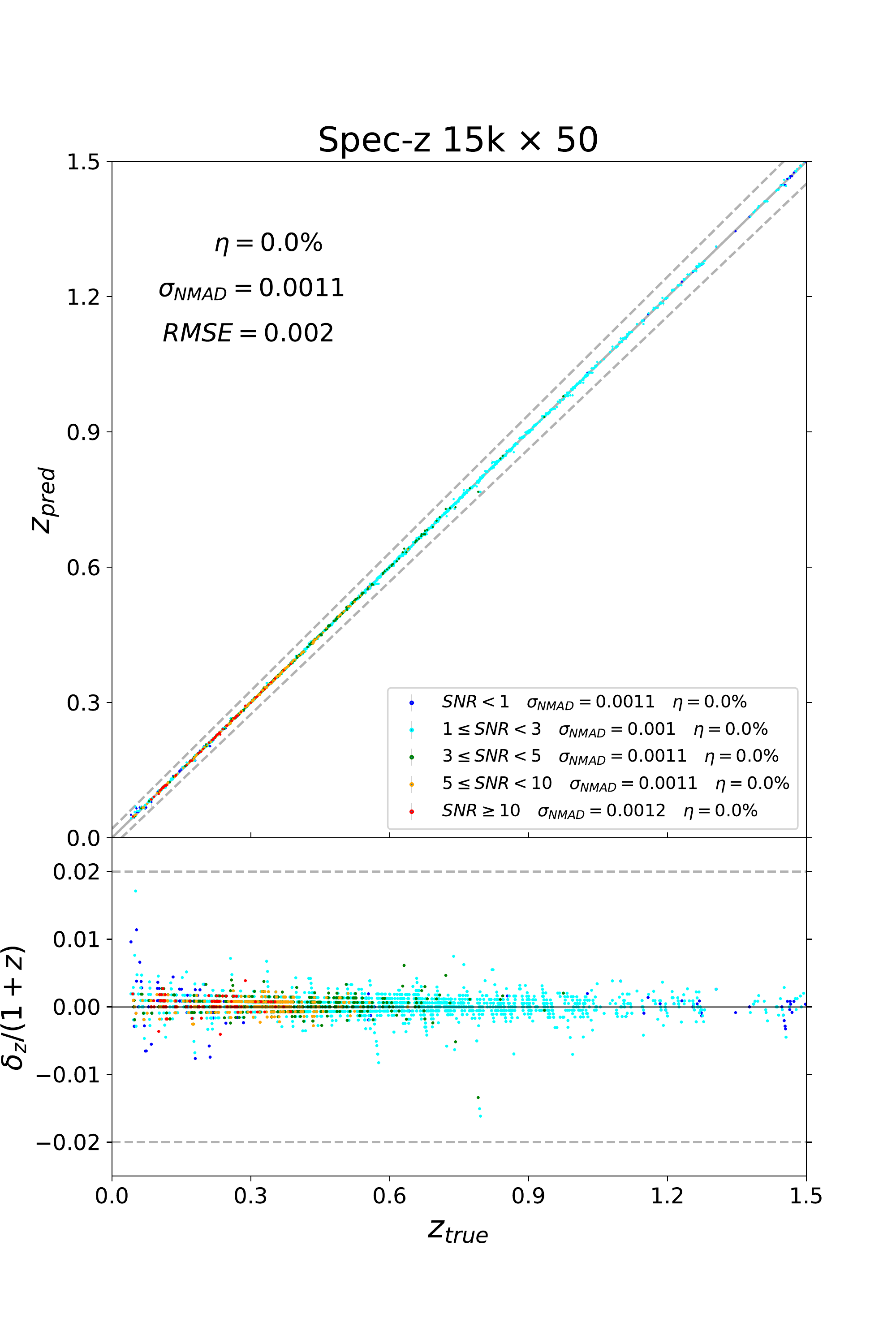}
			\end{minipage}
		}
	\caption{The best-fit spec-$z$ and 1-$\sigma$ errors $\sigma_z$ derived from the spec-$z$ PDFs for the testing data, which is estimated by the 1-d CNN for different number of training data, i.e. 50 Gaussian random realizations of 4,000, 6,000, 8,000, 10,000, 12,000, and 15,000 original spectral data. The results derived from the data with different total SNRs are also shown as blue (SNR$<1$), cyan ($1\le{\rm SNR}<3$), green ($3\le{\rm SNR}<5$), orange ($5\le{\rm SNR}<10$), and red (SNR$\ge10$) dots. The redshift outlier is denoted by dashed lines. We can find that the 1-d CNN can provide excellent results for the spectral of the CSST spectroscopic survey with $\sigma_{\rm NMAD}\simeq0.001$, as well as the data with low SNRs. More training data tend to get better accuracy and smaller outlier fraction $\eta$, and it seems to be saturate when the number of training data is greater than $\sim$10,000$\times$50.}
	\label{fig:spec_z_fitting}	
	\end{figure*}

	\begin{figure*}[ht]
		\centering
		\subfigure{
			\begin{minipage}[t]{0.31\linewidth}
				\centering
				\includegraphics[scale=0.31]{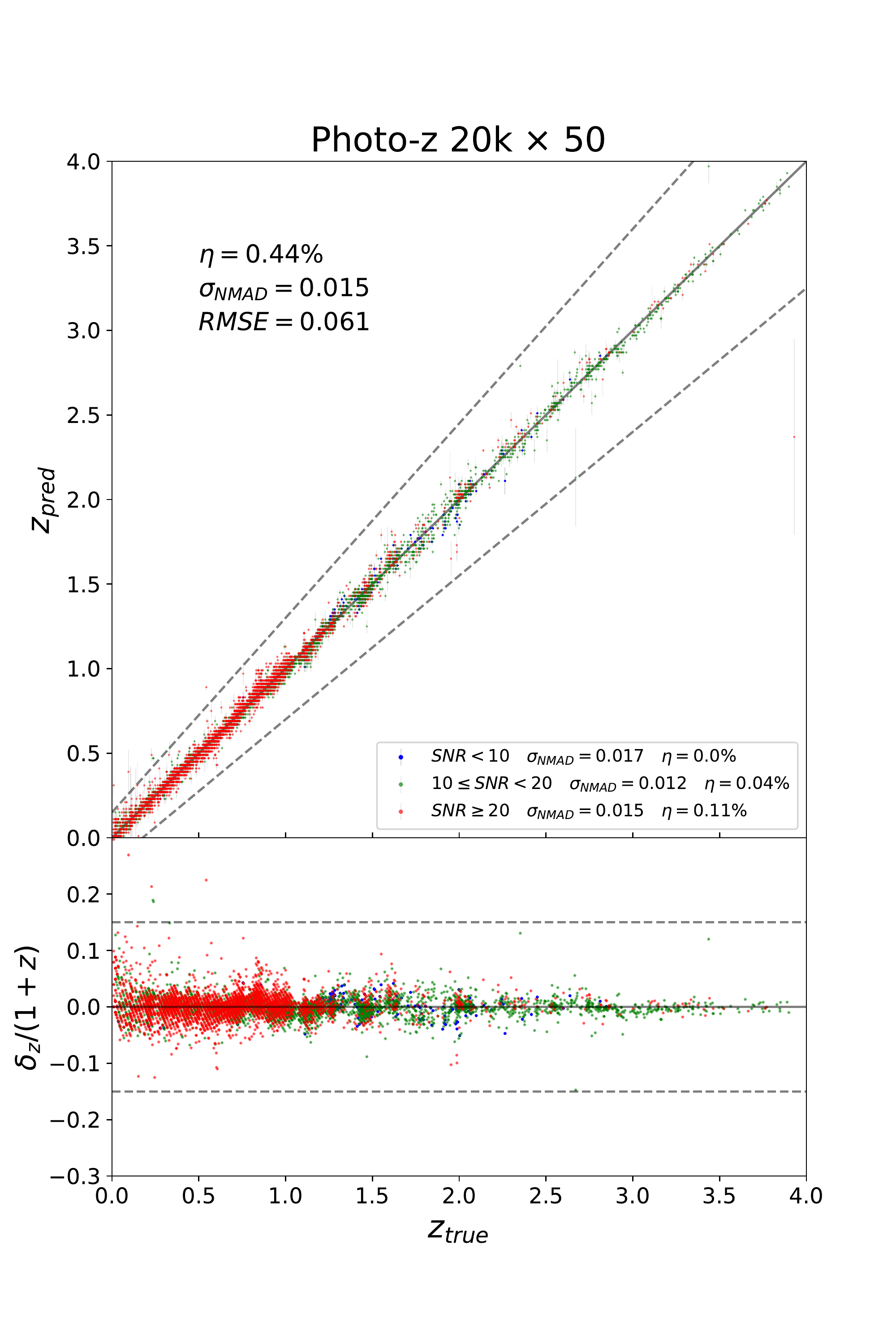}
			\end{minipage}
		}
		\subfigure{
			\begin{minipage}[t]{0.31\linewidth}
				\centering
				\includegraphics[scale=0.31]{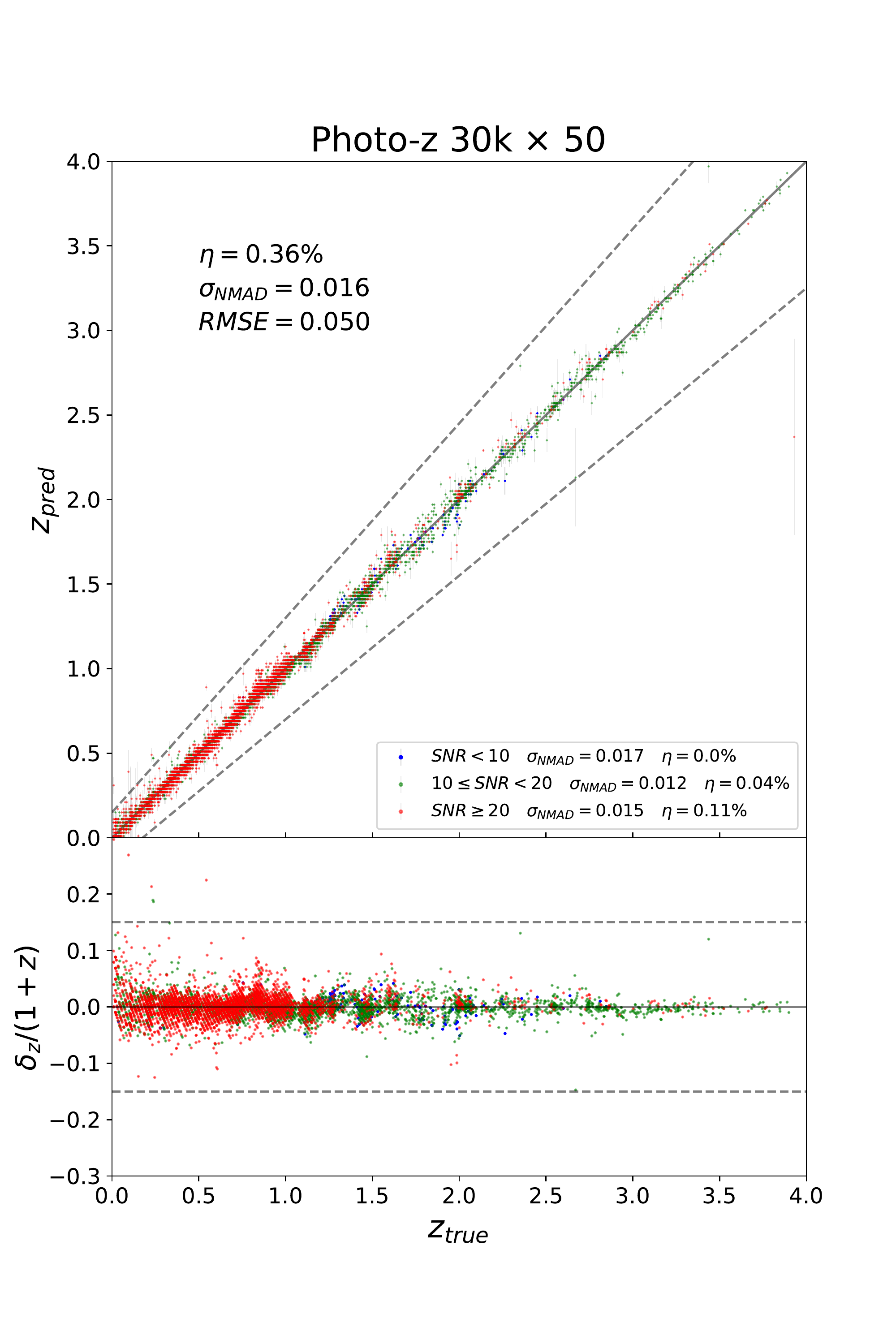}
			\end{minipage}
		}
		\subfigure{
			\begin{minipage}[t]{0.31\linewidth}
				\centering
				\includegraphics[scale=0.31]{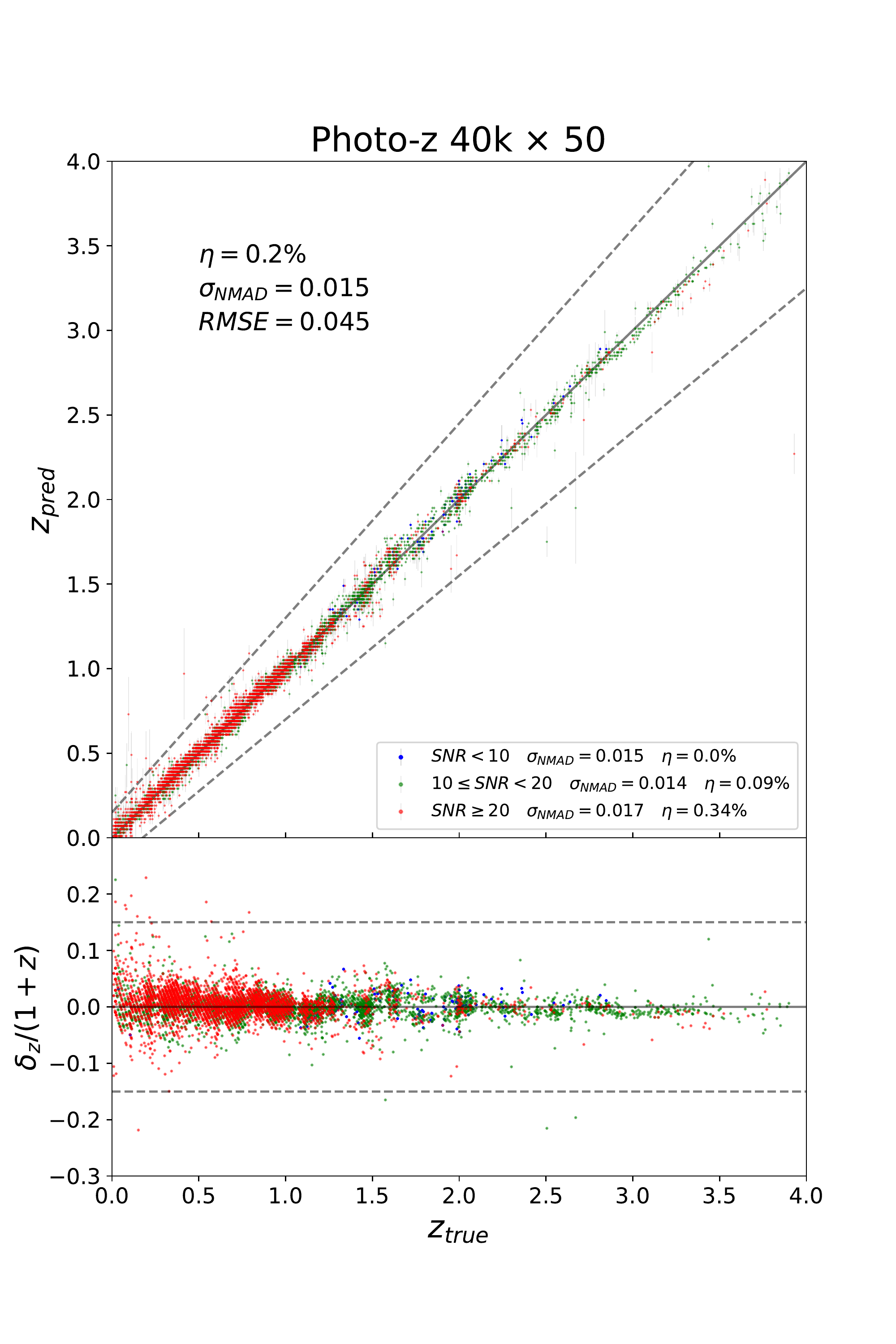}
			\end{minipage}
		}
		
		\subfigure{
			\begin{minipage}[t]{0.31\linewidth}
				\centering
				\includegraphics[scale=0.31]{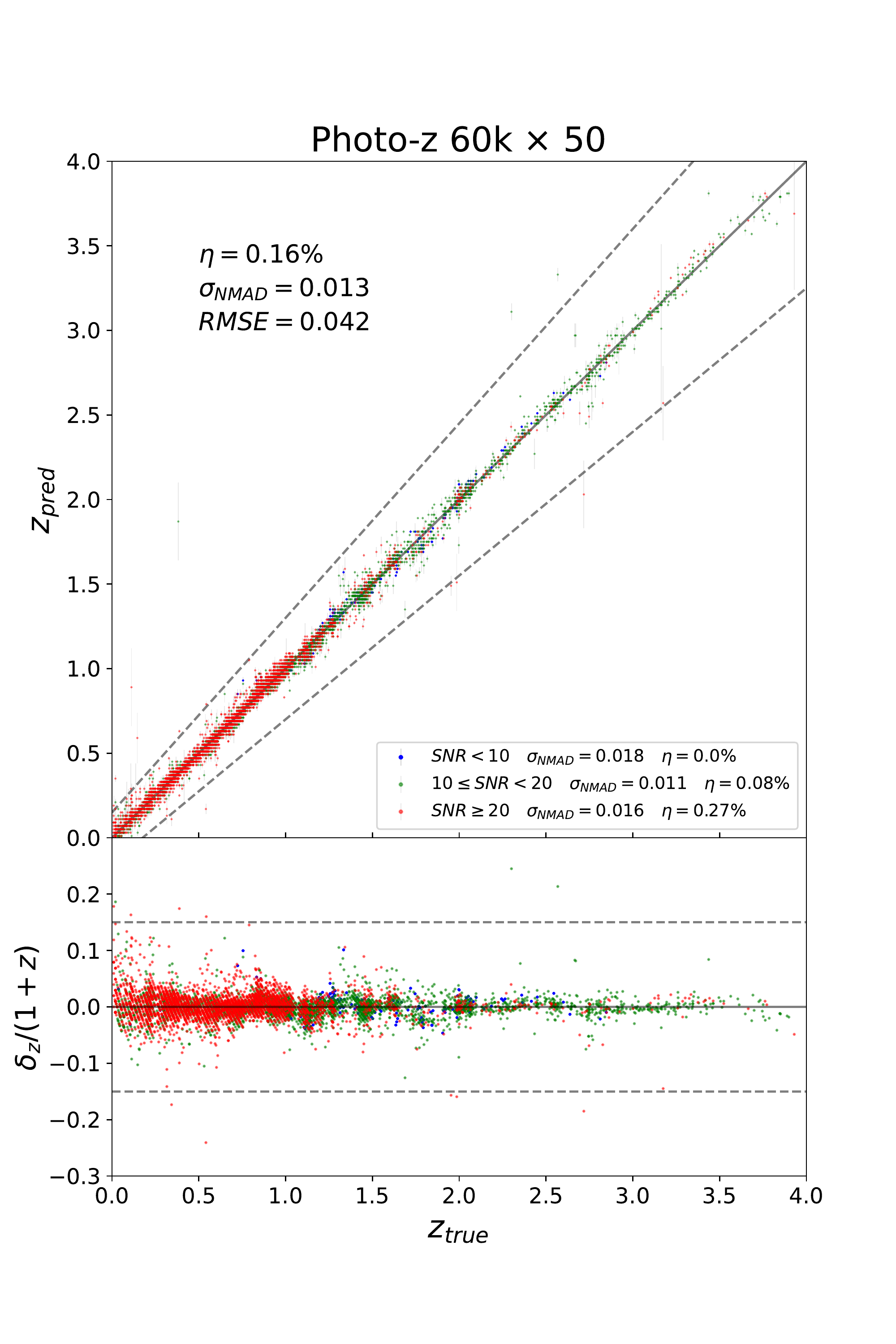}
			\end{minipage}
		}
		\subfigure{
			\begin{minipage}[t]{0.31\linewidth}
				\centering
				\includegraphics[scale=0.31]{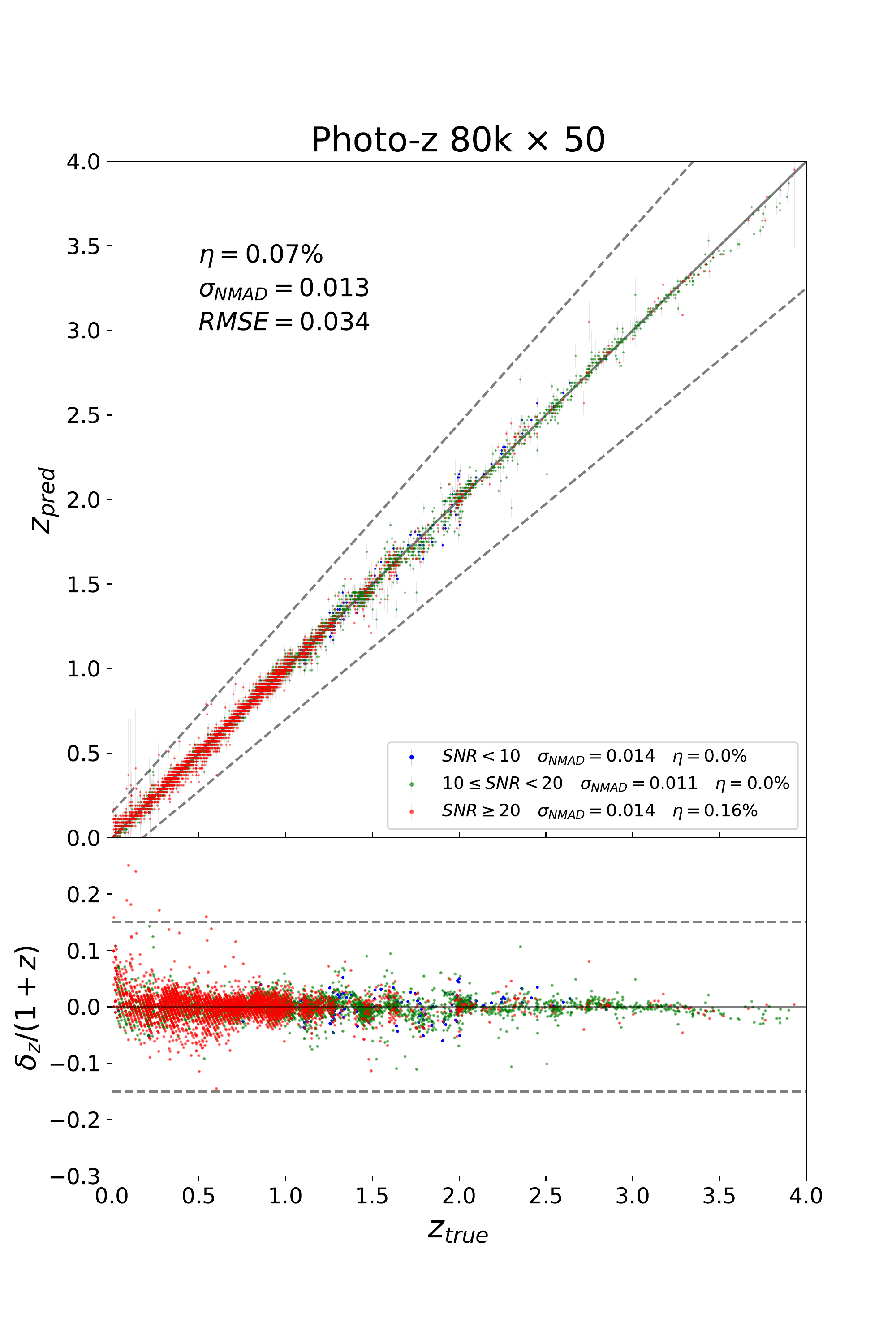}
			\end{minipage}
		}
		\subfigure{
			\begin{minipage}[t]{0.31\linewidth}
				\centering
				\includegraphics[scale=0.31]{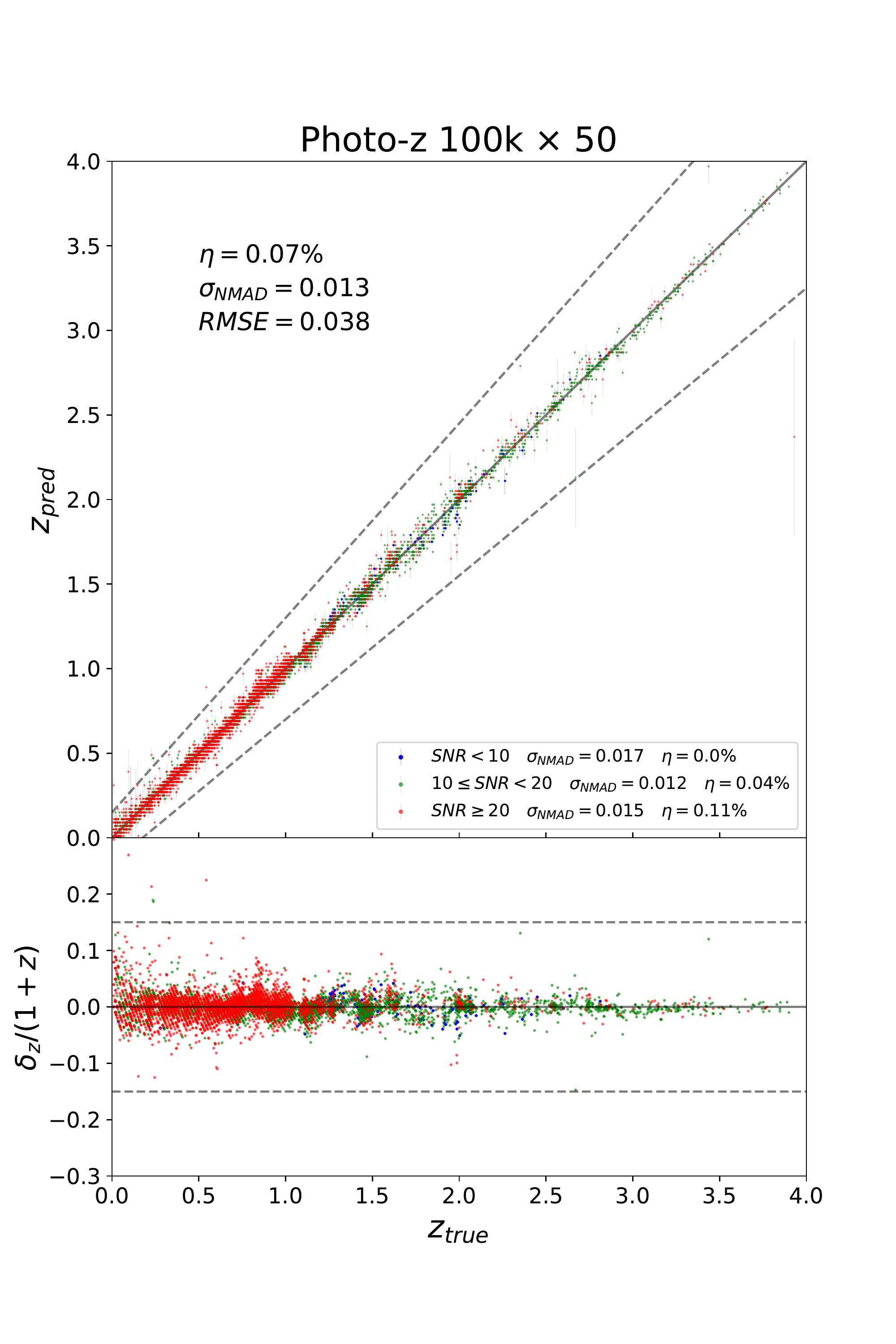}
			\end{minipage}
		}
		\caption{The best-fit photo-$z$ and 1-$\sigma$ errors $\sigma_z$ derived from the photo-$z$ PDFs for the photometric testing data, which is estimated by the MLP for different numbers of training datasets, i.e. 50 Gaussian random realizations of 20,000, 30,000, 40,000, 60,000, 80,000, and 100,000 original photometric data. The results derived from the data with different SNRs are also shown as blue (SNR$<10$), green ($10\le{\rm SNR}<20$), and red (SNR$\ge20$) dots. Note that the SNR is only for $i$ band here. The redshift outlier is denoted by dashed lines. We can find that the MLP we use can provide accurate photo-$z$ estimates with $\sigma_{\rm NMAD}\simeq0.01$, which is smaller than the SED template-fitting method by a factor of two at least. Besides, similar as the spec-$z$ case, the results of the low-SNR data has comparable accuracy to the high-SNR data, and basically more training data can obtain higher accuracy and smaller outlier fraction $\eta$ when the training data number is less than $\sim$80,000$\times50$.}
		\label{fig:photo_z_fitting}
	\end{figure*}
	
\begin{figure*}
\centering
\includegraphics[width=3.5in]{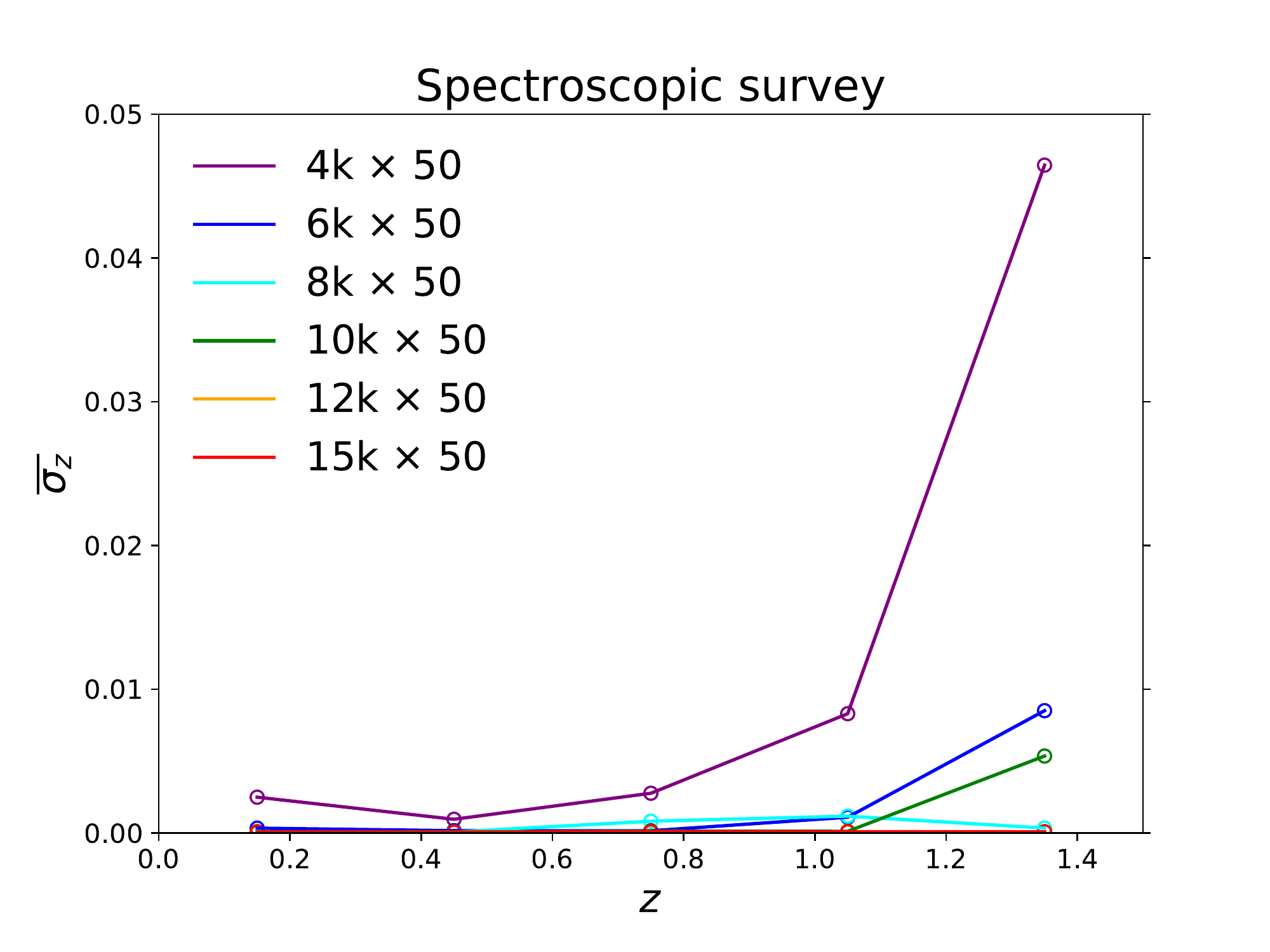}
\includegraphics[width=3.5in]{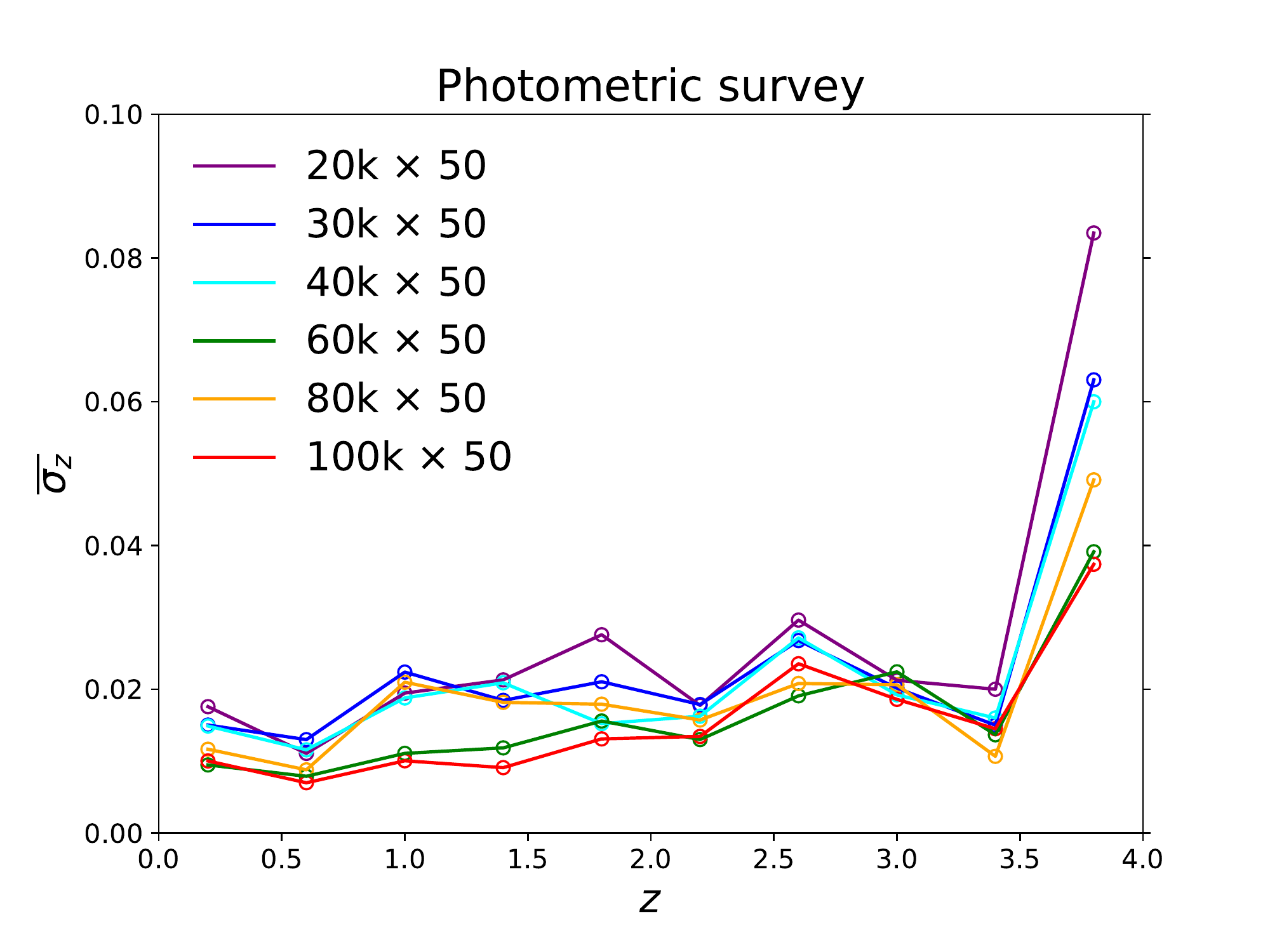}
\caption{The mean redshift errors $\overline{\sigma_z}$ as a function of $z$ for spec-$z$ (left panel) and photo-$z$ (right panel) analysis. The redshift errors are averaged over redshift bins $\Delta z=0.3$ and 0.4 for the spec-$z$ and photo-$z$ surveys, respectively. The results from different numbers of training data have been shown. We can find that $\overline{\sigma_z}$ is basically stable and low at $z<1$ and 3.5 for spec-$z$ and photo-$z$, respectively, and increases quickly at higher redshifts.}
\label{fig:z_error}
\end{figure*}

\section{Result}\label{sec:Result}

In order to assess the accuracy of derived spec-$z$ and photo-$z$, we define the catastrophic redshift or outlier as $|\Delta z|/(1+z_{\rm true})>D_{\rm out}$, where $\Delta z=z_{\rm pred}-z_{\rm true}$, and $z_{\rm pred}$ and $z_{\rm true}$ are the predicted and true redshifts, respectively. $D_{\rm out}=0.02$ and 0.15 for spec-$z$ and photo-$z$ analysis, respectively. The outlier fraction is denoted as $\eta$ here. We also adopt the normalized median absolute deviation \citep[NMAD,][]{Brammer08}, and it is given by
\be
\sigma_{\rm NMAD} = 1.48 \times {\rm median}\left(  \left|  \frac{\Delta z-{\rm median}(\Delta z)}{1+z_{\rm true}} \right|\right).
\ee
This definition has advantages that it can naturally suppress the weights of catastrophic redshifts, and provide a proper estimate of the redshift accuracy. Besides, the PDFs, best-fits, and 1-$\sigma$ errors of the predicted redshifts are also derived, based on the uncertainties of the testing data. We randomly generate 50 samples from a Gaussian distribution based on the flux and error at a wavelength for each testing data, and input them into the networks to get 50 predicted redshifts. Then the PDF, best-fit (the value at the PDF peak) and 1-$\sigma$ error of the redshift can be derived from the number distribution of these redshift points. Examples of photo-$z$ PDFs for different $i$-band SNRs are shown in Figure~\ref{fig:photo_z_pdf}. The PDFs of spec-$z$ are quite narrow like spikes with width d$z$=0.002. We find that 50 random samples are good enough to produce the PDFs, more samples would not change the result significantly. Note that we don't consider the uncertainty from the neural network, since we have chosen the best network model, which gives the most accurate redshift estimate, from a set of well-trained networks, and can effectively reduce this effect from the network\footnote{After performing a few tests, we find that the uncertainty from the networks is always smaller than that from data, and it would not affect the results if the networks are well trained.}.

In Figure~\ref{fig:spec_z_fitting}, we show the best-fit spec-$z$ results of the spectroscopic testing data for different numbers of training sample, which are randomly selected from the whole training dataset. The 1-$\sigma$ errors $\sigma_z$ of the predicted spec-$z$ are shown in gray bars. Both the best-fits and errors are derived from the spec-$z$ PDFs obtained by the network. The root mean square error (RMSE) for the whole sample is also calculated to further indicate the effect of outlier. The results of the spectroscopic data with different SNR ranges are also shown and denoted by different colors. The fractions of the data with total SNR$<1$, $1\le{\rm SNR}<3$, $3\le{\rm SNR}<5$, $5\le{\rm SNR}<10$, and SNR$\ge10$ are 6.3\%, 71.2\%, 13.0\%, 7.6\%, and 1.9\%, respectively. As we can see, most of the CSST slitless spectral data have SNRs less than 3, which is challenging for redshift derivation. However, we find that the redshifts obtained from the low SNRs still have high accuracy with $\sigma_{\rm NMAD}\simeq0.001$, which is comparable to the high SNR data. This is because we have sufficiently considered the statistical effects resulting from the data with large noises, and generated the corresponding noisy data to train the networks. This indicates that the neural network has advantages to derive the spec-$z$ from low SNRs data, as long as we can generate sufficient training sample and train it with real-enough noisy spectra. 

We also find that more training data can give better results with smaller outliers, and the results will be good enough and would not notably change when the number of training data greater than 10,000 with 50 Gaussian random realizations. We also test the case that no errors are inputed into the 1-d CNN for training, and find that $\sigma_{\rm NMAD}=0.002$ and outlier fraction $\eta=10.12\%$ for 12,000$\times$50 case, which are much larger than that with errors included. This implies that the errors contain useful information that can be helpful for network training to derive more accurate redshift.

In Figure~\ref{fig:photo_z_fitting}, the best-fit photo-$z$ and 1-$\sigma$ errors $\sigma_z$ of the photometric testing data extracted from the MLP for different numbers of training data have been shown. The photo-$z$ best-fits and errors are estimated from the photo-$z$ PDFs derived by the network. The results of the data with different SNR ranges are also listed. The percentages of the data with SNR$<10$, $10\le{\rm SNR}<20$, SNR$\ge20$ are found to be 2.4\%, 53.0\% and 44.6\% respectively. Note that the SNR is only for $i$ band here, it could be less than 10 since we select the data with SNR$\ge10$ for $i$ or $g$ band. We can see that the MLP can provide good predications on photo-$z$, and the $\sigma_{\rm NMAD}$ is close to 0.01 for the number of training data greater than 60,000$\times$50. In addition, the fraction of catastrophic redshift or outlier is quite low with $\eta<0.5\%$ for all cases we explore. The RMSE is also given to show the effect of outlier. We find that our MLP can provide similar accuracy for photo-$z$ estimation compared to exiting codes \citep[e.g.][]{Collister04, Gerdes10, Samui17}. By comparing to the results from the SED templating-fitting method given by \cite{Cao18}, we can find that our results of $\sigma_{\rm NMAD}$ are improved by a factor of two at least, and the outlier is dramatically suppressed by more than one order of magnitude. Besides, like the results of the spec-$z$, the photo-$z$ accuracy from the low-SNR data is similar as the high-SNR data, which proves that our method of generating the training data is feasible and effective. If not including errors in the training process, we obtain $\sigma_{\rm NMAD}=0.053$ and $\eta=9.09\%$ for 80,000$\times$50 case, which proves again that the errors can be helpful to improve the redshift estimate accuracy in the training.

In Figure~\ref{fig:z_error}, we show the mean redshift errors $\overline{\sigma_z}$ (the mean 1-$\sigma$ error derived from the PDFs for all galaxies at $z$) as a function of redshift for different numbers of training data for the CSST spectroscopic and photometric surveys. We find that the mean errors are lower than 0.002 (except for 4,000$\times$50 case) at $z\lesssim 1$ and 0.03 at $z\lesssim 3.5$ for spec-$z$ and photo-$z$, respectively. At higher redshifts, $\overline{\sigma_z}$ increases quickly, due to fewer training data are generated and used in these ranges. The spec-$z$ accuracy at $1<z<1.5$ actually can be further increased if more high-quality spectra data can be used from ongoing or future high-resolution spectroscopic surveys.

\section{Summary}\label{sec:Summary}

We explore the accuracies of spec-$z$ and photo-$z$ that can be derived from the CSST slitless spectroscopic and photometric surveys by neural networks. The 1-d CNN and MLP are adopted to extract spec-$z$ and photo-$z$, respectively. We assume the high-quality galaxy spectra and redshifts can be obtained by ongoing and future powerful spectroscopic surveys, which can be used as training data. We generate the mock data for the CSST spectroscopic and photometric surveys based on the COSMOS catalog, considering the CSST observational and instrumental effects. About 20,000 and 126,000 galaxies are selected for the CSST spectroscopic and photometric surveys, respectively.

The 1-d CNN and MLP we adopt are developed based on the Keras API. We have tested different architectures, and choose to use two convolutional and two fully connected layers for the 1-d CNN in the spec-$z$ analysis, and two hidden layers for the MLP in photo-$z$ estimate. The mock data are divided into training and testing datasets. About 4,500 and 10,000 data are randomly selected as the testing data for the spec-$z$ and photo-$z$ analysis, respectively, and the rests are as training datasets. In order to reduce the statistical effect on the observational data and generate enough training sample for the same galaxy type, we create 50 Gaussian random realizations of the original training data to train the networks. More realizations would not change the results significantly. After validation process, we find that our networks can be well trained to get accurate redshift prediction.

We also try to derive the errors of predicted redshifts by generating random samples based on the testing data. The PDFs can be properly illustrated by about 50 realizations, and then the best-fits and 1-$\sigma$ errors can be obtained. We find that the neural network can provide excellent estimates on both spec-$z$ and photo-$z$, which is significantly better than the ordinary SED template-fitting method, especially for noisy data. The redshift accuracy can achieve 0.001 and 0.01 for spec-$z$ and photo-$z$, respectively, with extremely small outlier fractions. Our MLP can reach similar photo-$z$ accuracy with more efficient training process compared to existing codes. In particular, by using our method, the results of low-SNR data are comparable to high-SNR data, that indicates the advantages of neural network in future redshift analysis.

\acknowledgments

We thank Fengshan Liu and the CSST instrument team for helpful discussions. XCZ and YG acknowledges the support of NSFC-11822305, NSFC-11773031, NSFC-11633004, MOST-2018YFE0120800, the Chinese Academy of Sciences (CAS) Strategic Priority Research Program XDA15020200, the NSFC-ISF joint research program No. 11761141012, and CAS Interdisciplinary Innovation Team. XZ acknowledges the support of National Natural Science Foundation of China (Grant No. U2031143). XLC acknowledges the support of CAS QYZDJ-SSW-SLH017. ZHF acknowledges the support of NSFC-11333001. LPF acknowledges the support from NSFC grants 11722326, 11673018 \& 11933002, STCSM grants 18590780100, 19590780100 \& 188014066, SMEC Innovation Program 2019-01-07-00-02-E00032 \& Shuguang Program 19SG41. This work is partially supported by the China Manned Space Program through its Space Application System.

\end{document}